\documentclass{aa}
\usepackage{graphicx}
\usepackage{txfonts}
\usepackage[colorlinks]{hyperref}
\hypersetup{colorlinks=true,linkcolor=blue,citecolor=blue,filecolor=blue,urlcolor=blue,}
\usepackage{amsmath}
\usepackage{comment}
\usepackage{algorithm}
\usepackage{amsfonts}
\usepackage{bm}
\usepackage{rotating}
\usepackage{color}
\usepackage{graphicx}
\usepackage{subfigure}
\usepackage{tabularx}
\usepackage{lineno}
\usepackage{savesym}
\usepackage[flushleft]{threeparttable}
\savesymbol{tablenum}
\usepackage{siunitx}
\restoresymbol{SIX}{tablenum}

\def \a3 {A$^3$COSMOS }

\begin{document}

   \title{The observed total star formation rate function up to $z \sim 6$: complementary UV and IR contributions and comparison with state-of-the-art galaxy formation models}

   \author{A. Traina \inst{1}, C. Gruppioni \inst{1}, I. Delvecchio \inst{1},  B. Magnelli \inst{2}, F. Calura \inst{1}, L. Bisigello \inst{3}, A. Feltre \inst{4}, L. Vallini \inst{1}, G. De Lucia \inst{5,6}, F. Fontanot \inst{5,6}, M. Hirschmann \inst{7,5}, A. Katsianis \inst{8}, M. Parente \inst{5,6}, O. Cucciati \inst{1} L. Xie \inst{9} \ E. Schinnerer \inst{10}, D. Liu \inst{11}, S. Adscheid \inst{12}, H. S. B. Algera \inst{13}, M. Behiri \inst{14,15}, F. Gentile \inst{2}, S. Gillman \inst{16,17}, F. Pozzi \inst{18,1}, G. Zamorani \inst{1}
          } 

   \institute{Istituto Nazionale di Astrofisica (INAF) - Osservatorio di Astrofisica e Scienza dello Spazio (OAS), via Gobetti 101, I-40129 Bologna, Italy
         \and
             Université Paris-Saclay, Université Paris  Cité, CEA, CNRS, AIM, F-91191 Gif-sur-Yvette, France
         \and     
             INAF - Osservatorio Astronomico di Padova, Vicolo dell'Osservatorio, 3, I-35122, Padova, Italy
         \and 
             INAF - Osservatorio Astrofisico di Arcetri, Largo E. Fermi 5, 50125, Firenze, Italy
         \and    
             INAF – Astronomical Observatory of Trieste, via G.B. Tiepolo 11, I-34143 Trieste, Italy
         \and
             IFPU - Institute for Fundamental Physics of the Universe, via Beirut 2, 34151, Trieste, Italy
         \and
            Institute for Physics, Laboratory for Galaxy Evolution and Spectral Modelling, Ecole Polytechnique Federale de Lausanne, Observatoire de Sauverny, Chemin Pegasi 51, CH-1290 Versoix, Switzerland
         \and   
            School of Physics and Astronomy, Sun Yat-sen University, Zhuhai Campus, 2 Daxue Road, Xiangzhou District, Zhuhai, China  
         \and
            Tianjin Normal University, Binshuixidao 393, Tianjin, China             
         \and  
            Max Planck Institut für Astronomie, Königstuhl 17, D-69117 Heidelberg, Germany
         \and    
            Purple Mountain Observatory, Chinese Academy of Sciences, 10 Yuanhua Road, Nanjing 210023, China
         \and   
            Argelander-Institut für Astronomie, Universität Bonn, Auf dem Hügel 71, 53121 Bonn, Germany
         \and
            Institute of Astronomy and Astrophysics, Academia Sinica, 11F of Astronomy-Mathematics Building, No. 1, Section 4, Roosevelt Road, Taipei 106216, Taiwan, R.O.C.
         \and 
             SISSA, Via Bonomea 265, 34136 Trieste, Italy
         \and 
             IRA-INAF, Via Gobetti 101, 40129 Bologna, Italy    
         \and
             Cosmic Dawn Center (DAWN), Denmark
         \and 
            DTU-Space, Technical University of Denmark, Elektrovej 327, DK-2800 Kgs. Lyngby, Denmark
         \and             
             Dipartimento di Fisica e Astronomia (DIFA), Università di Bologna, via Gobetti 93/2, I-40129 Bologna, Italy
             }

   \date{Received ??; accepted ??}

  \abstract
   {}
   {We investigate how the obscured IR-derived and the dust-corrected UV star formation rate functions (SFRFs) compare with each other, and with predictions from state-of-the-art theoretical models of galaxy formation and evolution.}
   {We derive the IR-SFRF from the ALMA A$^3$COSMOS survey, by converting the IR luminosity functions (IR-LFs) into SFRF after correcting for AGN contribution. Similarly, we obtain the UV SFRFs from literature UV LFs, corrected for dust-extinction. First, we fit the two SFRFs independently via a MCMC approach, then we combine them to obtain the first estimate of the “total” SFRF out to $z \sim 6$. Finally, we compare this SFRF with the predictions of a set of theoretical models.}
    {We derived the UV (dust-extinction corrected, from literature UV-LFs) and IR SFRFs (from \textit{Herschel} and ALMA IR-LFs) at $0.5 < z < 6$ , finding that they are mostly complementary, covering different ranges in star formation rate (SFR$ < 10-100$ M$_{\odot}$yr$^{-1}$ for the UV-corrected and SFR$ > 100$ M$_{\odot}$yr$^{-1}$ for the IR). From the comparison of the total SFRF with model predictions we find an overall good agreement at $z < 2.5$, with increasing difference at higher redshifts, with all models missing the galaxies that are forming stars with the highest SFRs. We finally obtained the UV (dust-corrected), IR and total star formation rate densities (SFRDs), finding that there are no redshift ranges where UV and IR alone are able to reproduce the whole total SFRD.} 
   {}


    \titlerunning{SFRF vs simulations}
    \authorrunning{A. Traina et al.}
    \maketitle
    
%

\section{Introduction}\label{sec:intro}

Understanding how galaxies form and evolve through cosmic time has been a topic of interest of both observational and numerical astrophysics for the past decades up to present day. The characterization of all the physical mechanisms behind the evolution of galaxies is not an easy task and requires a variety of different observation strategies (i.e., photometric as well as spectroscopic). Numerical simulations and theoretical models are tools that have been widely used in the community to investigate the physics that governs galaxy formation and evolution \citep{white1991sam,Kauffmann1993sam,springel2001sam,bower2006sam,croton2006sam,monaco2007sam,somerville2008sam,fontanot2009sam,guo2011sam,benson2012sam,menci2012sam,somerville2012sam,henriques2013sam,gruppioni2015sam,henriques2015sam,schaye2015eagle,pillepich2018illustrisTNG,dave2019simba,lacey2016sam,lagos2018sam}. Observational studies have shown that star-forming galaxies (SFGs) populate a relatively tight “main sequence” (MS), relating stellar mass and star formation rate \citep[][]{noeske2007ApJ...660L..43N,Whitaker2012ApJ...754L..29W,speagle2014MS}. The small scatter of the MS suggests that galaxies grow in a quasi-equilibrium state, regulated by the balance between gas accretion, star formation, and outflows \citep[][]{bouche2010ApJ...718.1001B,lilly2013ApJ...772..119L,dave2012MNRAS.421...98D}. In the $\Lambda$CDM framework, galaxies can be fueled by cold gas inflows from the cosmic web \citep[][]{keres2005MNRAS.363....2K,dekel2009Natur.457..451D}, while stellar and AGN-driven outflows regulate their efficiency at converting gas into stars \citep[][]{hopkins2014MNRAS.445..581H,somerville_dave2015ARA&A..53...51S,naab_ostriker2017ARA&A..55...59N}. Both processes are incorporated into theoretical models through sub-grid prescriptions in simulations or empirical recipes in SAMs, where they strongly shape the predicted star formation rate function (SFRF) \citep[][]{dubois2016MNRAS.463.3948D}. Comparing UV+IR observations with such models is therefore crucial for assessing the fidelity of feedback and dust implementations \citep[][]{picouet2023A&A...675A.164P}.
To trace the evolution of a galaxy, one of the key aspects is to investigate the assembly of its mass and time scale on which the gas is converted into newly formed stars. In order to probe how this occurs, the star formation rate (SFR) represents the most accurate indicator for retrieving information on the instantaneous growth in stellar mass ($M_{\star}$).
\par Although the SFR of galaxies is an informative parameter to investigate galaxy evolution, its derivation is subject to several limitations, depending on the wavelengths of observation. Different methodologies involve either detailed spectral energy distribution (SED) fitting or the use of star formation indicators. However, measurements derived from luminosities in different bands like the ultraviolet (UV) or the infrared (IR), as well as individual nebular emission lines (e.g., H${\alpha}$) can lead to diverging estimates for the SFR. Each of these probes of the luminosity of a galaxy have been calibrated in the local Universe to a corresponding SFR \citep[e.g.,][]{kennicutt1998sfr,kennicutt1998sfruv,kennicutt2012sfr,smit2012sfruv} and it is unclear if the same calibrations and assumptions (e.g. IMF) can be used for higher redshifts. In particular, UV and IR emission trace SF on very similar timescales \citep[see][]{kennicutt2012sfr}.
From a theoretical perspective, UV and IR luminosities of simulated galaxies are often derived by applying dust radiative transfer post-processing to the intrinsic stellar populations \citep[e.g.,][]{trayford2017MNRAS.470..771T,narayanan2021ApJS..252...12N}. The SIMBA simulation represents an exception, as it includes a self-consistent dust production model \citep[][]{dave2019simba}, enabling direct predictions of both obscured and unobscured star formation. As shown by \citet{picouet2023A&A...675A.164P}, the choice of feedback implementation and dust modeling critically affects the ability of models to reproduce the observed UV- and IR-derived SFRFs.
\par Observations in the optical and UV have enabled estimates of the star formation rate density (SFRD) up to redshifts of $z \sim 7 - 8$ \citep[see, e.g.,][]{bouwens2014sfrd, oesch2015sfrd, laporte2016sfrd, oesch2018sfrd}, and even as high as $z \sim 10$ \citep[e.g.,][]{harikane2023jwst}, significantly extending our understanding of star formation in the early universe. However, these measurements—primarily based on rest-frame UV data—are likely missing the most dust-obscured and thus star-forming galaxies. On the other hand, IR LFs are still poorly constrained beyond $z \sim 5$, with the faint-end in particular remaining largely unconstrained \citep[even though some attempts have been made to estimate it, see e.g.,][]{barrufet2023, fujimoto2024sfrd}.
\par In the last fifteen years, many works studied the main processes regulating the SF and have compared the observed star formation rate function (SFRF) \citep[][]{reddy2008sfrf,oesch2010sfrf,vander2010sfrf,ly2011sfrf,magnelli2011sfrf,cucciati2012sfrd,smit2012sfruv,gruppioni2013lf,magnelli2013ir,patel2013sfrf,sobral2013sfrf,duncan2014sfrf,bouwens2015sfrf,alavi2016sfrf,parsa2016sfrf,gruppioni2020alpine} with predictions from simulations and models \citep{dave2011simul,fontanot2012sam,tescari2014simul,gruppioni2015sam,katsianis2017eagle,katsianis2021sfrd}. 
These studies reveal tensions between observations and models, often linked to feedback implementations and dust treatments. While UV and IR estimates provide complementary information, models without adequately calibrated AGN/SN feedback or dust prescriptions cannot simultaneously reproduce both the faint and bright ends of the SFRF \citep[][]{gruppioni2015sam, picouet2023A&A...675A.164P}.
\par In order to investigate the evolution of the star formation rate with the redshift, and to alleviate the aforementioned tensions, large surveys are needed. The automated mining of the ALMA Archive in COSMOS \citep[A$^3$COSMOS,][]{liu2019a31,liu2019a32}, which is the largest ALMA survey to date, represents an ideal benchmark for simulations and semi-analytical models (SAMs), probing the IR-mm SFRF over a wide redshift range ($z \sim 0-6$). A physical and statistical analysis of the \a3 galaxy sample has already been performed in a previous work by \cite{traina2024sfrd}, hereafter T24, aimed at deriving the IR luminosity function (IR-LF) and cosmic dust-obscured star formation rate density (SFRD) at different redshifts. From the same database, with the addition of the GOODS-S ALMA archival images, \citet{adscheid2024a3cosmos} derived the (sub-)mm number counts, while other physical properties (e.g., molecular gas, dust attenuation) have been explored in previous works by \citet{liu2019a32,fudamoto2020dust_att, wang2022molgas}.
In this paper, we use the IR-LFs previously derived in T24 from the A$^3$COSMOS along with UV-LFs from the literature to obtain new estimates of the combined\footnote{In this paper, we refer to the “combined” SFRF and SFRD as that obtained by combining SFRF data points from both bands, which sample different intervals in SFR.} (IR and UV) SFRF at $0< z < 6$ and compare it with predictions from the literature from state-of-the-art hydrodynamical simulation and SAMs.

\par The paper is organized as follows: in Section \ref{sec:sample} we present the archival samples used to derive the SFRFs; in Section \ref{SFRF} we derive the IR-SFRF for the \a3 survey, combine it to observational results from the UV and compare to fiducial models prediction of the SFRF from a set of hydrodynamical simulation and SAMs in Section \ref{sec:comparison_w_models}; in Section \ref{sfrd} we show the comparison between observed SFRD and the predicted one; in Section \ref{endings} the conclusions are presented.
Throughout the paper, we assume a \cite{chabrier2003imf} stellar
initial mass function (IMF) and adopt a $\Lambda$CDM cosmology with $H_{0} = 70$ $\rm km$ $\rm s^{-1}$ $\rm Mpc^{-1}$, $\Omega_{\rm m} = 0.3$, and $\Omega_{\Lambda} = 0.7$.

%
%
\section{The data}\label{sec:sample} 

\subsection{IR-mm Sample}

\begin{figure}[]
\centering
{\includegraphics[width=.48\textwidth]{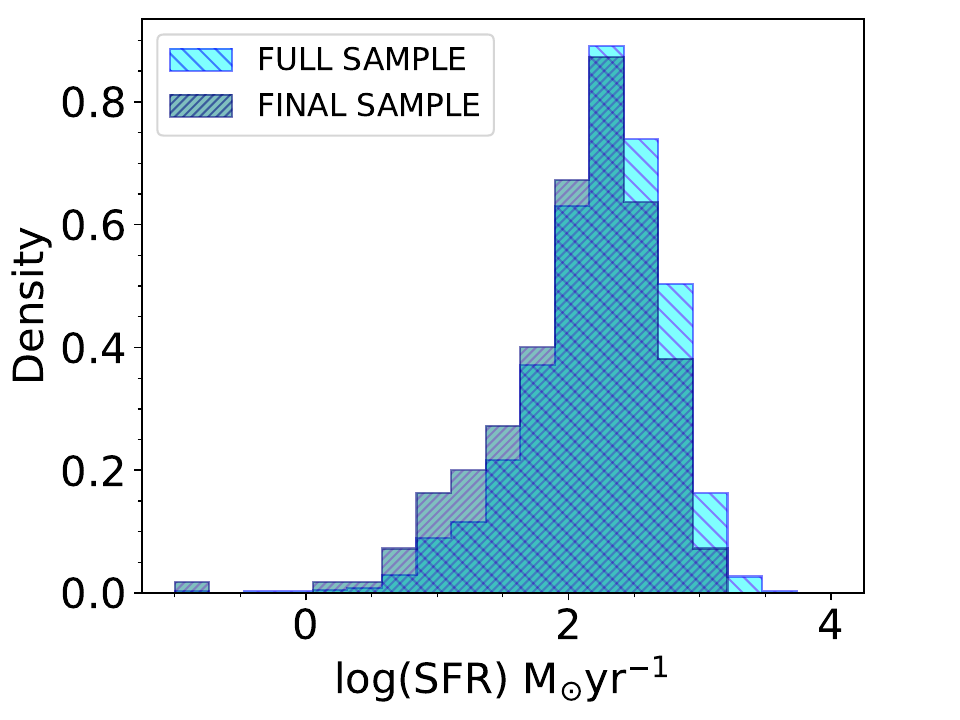}}
\caption{\small{Density (normalized to 1) distributions of the IR-SFR obtained through SED fitting for total A3COSMOS sample (cyan) and for the sub-sample considered in this work (teal).}}
\label{fig:sfr}
\end{figure}

The \a3 survey is the largest database of ALMA-selected galaxies. Indeed, it is the collection of all the archival ALMA observations within the COSMOS field \citep[][]{scoville2007cosmos,weaver2022cosmos2020}{}{}. In this paper, we used the most recent version of the \a3 combined with the COSMOS2020 \citep[][]{weaver2022cosmos2020}{}{} photometry \citep[][]{adscheid2024a3cosmos}{}{}. We use the IR-LF from T24, obtained by combining the \a3 database (turned into a “blind-like” survey) with \textit{Herschel} data points from \citet{gruppioni2013lf}. Below we summarize how the final \a3 sample is obtained and its main physichal properties. The sample in T24 is obtained as follows: for each ALMA pointing, we removed the sources within a radius of 1" from the center (which are identified as targets of the ALMA observations) to avoid selection bias, as well as those offset sources at a redshift similar to that of the target to correct for clustering bias. The sample is thus consisting of 189 (out of an initial sample of 1620) galaxies which are considered serendipitously detected in the ALMA pointings. The main integrated galaxy physical properties (i.e., stellar mass, dust luminosity, IR-SFR) were inferred in T24 through SED-fitting with the \texttt{CIGALE} SED fitting code \citep[][]{boquien2019cigale}{}{}, finding a massive ($M_{\star} \sim 10^{10}-10^{12}$ M$_{\odot}$), infrared luminous (L$_{IR,8-1000} \sim 10^{11}-10^{13}$ L$_{\odot}$) and highly star forming (SFR $\sim10-1000$ M$_{\odot}$yr$^{-1}$) population. Additionally, $\sim 40\%$ of the sources are hosting an AGN, with different AGN fractions (i.e., fraction of the AGN emission with respect to the total IR emission in the 5-40 $\mu$m range from the \texttt{CIGALE} templates, $f_{\rm AGN}$, ranging from 0.3 to 0.8). Figure \ref{fig:sfr} shows the SFR distribution of the full 1620 galaxies sample (cyan histogram) and for the subsample of 189 sources considered as serendipitous detection in T24 (teal histogram). Both distributions peak at SFR $\sim 300$ M$_{\odot}$yr$^{-1}$, but most objects were excluded from the parent A$^3$COSMOS sample because they are targets.

\subsection{UV datasets}\label{subsec:uv_datasets}
Although the star formation derived using the $L_{\rm IR}$ has been proven to account for a large fraction of the total SFR of a galaxy \citep[][]{wuyts2011sfr, pannella2015sfr}{}{}, it traces only the obscured part and the fractional contribution of the IR to UV derived SFRs may change at $z > 3$. For this reason, we included in our analysis the UV-LFs from seven works from the literature, broadly spanning the redshift range from 0 to 6 (to be consistent with our IR/mm sample): 
\begin{itemize}
    \item \cite{vander2010sfrf} derive the UV-LF in the range $3<z<5$, using $\sim 10^5$ Lyman-break galaxies from the CFHT Legacy Survey Deep fields.
    \item \cite{cucciati2012sfrd} study the UV-LF of $I$-band selected galaxies using the VVDS surveys, from the local Universe up to $z \sim 4.5$.
    \item \cite{parsa2016sfrf} explore the UV-LF from $z \sim 2$ to $z \sim 4$ combining the HUDF, CANDELS/GOODS-south and UltraVISTA/COSMOS surveys.
    \item \cite{mehta2017UV_LF} use the UVUDF to derive the UV-LF of Lyman-break galaxies at $z \sim 1.5 - 3$.
    \item \cite{ono2018uvLF} study the UV-LF towards higher redshifts ($z\sim 4 - 7$) through the Great Optically Luminous Dropout Research Using Subaru HSC (GOLDRUSH).
    \item \cite{adams2020uvlf} derive the LF at $z \sim 4$ combining the COSMOS survey and the \textit{XMM-Newton} Large-Scale Structure fileds.
    \item \cite{bouwens2021UV_LF,bouwens2022uvlf} use a combination of several fields to derive the UV-LF at $z \sim 2-9$.
\end{itemize}

Using these data, we take into account the unobscured contribution to the total SFRF at each redshift.

\section{The IR and UV SFR function}\label{SFRF}

To derive the SFRFs, we start from the IR and UV luminosity functions, converting $L_{\rm IR}$ and UV magnitudes into IR-SFR and UV-SFR, respectively. Specifically, for the conversion of $L_{\rm IR}$ to IR-SFR (which accounts for the obscured part of the SFR), we employed the \cite{kennicutt1998sfr} relation (for a Chabrier IMF):

\begin{equation}
SFR_{\rm IR} (M_{\odot} yr^{-1}) = 1.09 \times 10^{-10} \frac{L_{\rm IR}}{L_{\odot}}.
\end{equation}

  
To derive the component of the SFR from the UV luminosity, corrected for dust attenuation, we applied the following procedure. We corrected UV magnitudes in a self-similar manner via the same relation proposed by \citep[][]{hao2011sfr}:

\begin{equation}
M_{\rm UV, CORR} = M_{\rm UV, OBS} \times e^{\tau_{UV}} ,
\end{equation}

where $\tau_{UV}$ represents the effective optical depth ($\tau_{UV} = A_{1600} / 1.086$, with $A_{1600}$ being the dust absorption at 1600 \si{\angstrom}, calculated as $A_{1600} = 4.43 +1.99 \beta$). We adopt the form for $\beta$ as shown in \cite{bouwens2012sfr}:

\begin{equation}
\beta = \frac{d\beta}{dM_{\rm UV}} (M_{\rm UV,AB} + 19.5) + \beta_{M_{\rm UV} = 19.5},
\end{equation}

with $M_{\rm AB}$ being the absolute UV magnitude. For $\frac{d\beta}{dM_{\rm UV}}$ we assume the values tabulated by \cite{tacchella2013sfr} for various redshift intervals.

\par To derive the SFR from the UV luminosity, we used these UV-LF estimates corrected for dust attenuation (see Section \ref{total_sfrf} for a detailed motivation), converting the absolute UV magnitude (for the catalogs described in Section \ref{subsec:uv_datasets}) into UV luminosity:

\begin{equation}
    L_{\rm UV, corr} = 10^{-0.4 (M_{\rm UV, corr}-51.60)}
\end{equation}

Finally, from $L_{\rm UV, corr}$, we then calculated the UV corrected component of the SFR as follows \citep[see][]{kennicutt2012sfr}{}{}:

\begin{equation}
SFR_{\rm UV, corr} (\text{M}_{\odot} \text{yr}^{-1}) = 0.82 \times 10^{-28} \frac{L_{\rm UV, corr}}{\text{erg} \text{s}^{-1}  \text{Hz}^{-1}},
\end{equation}

\begin{figure*}[]
\centering
{\includegraphics[width=1\textwidth]{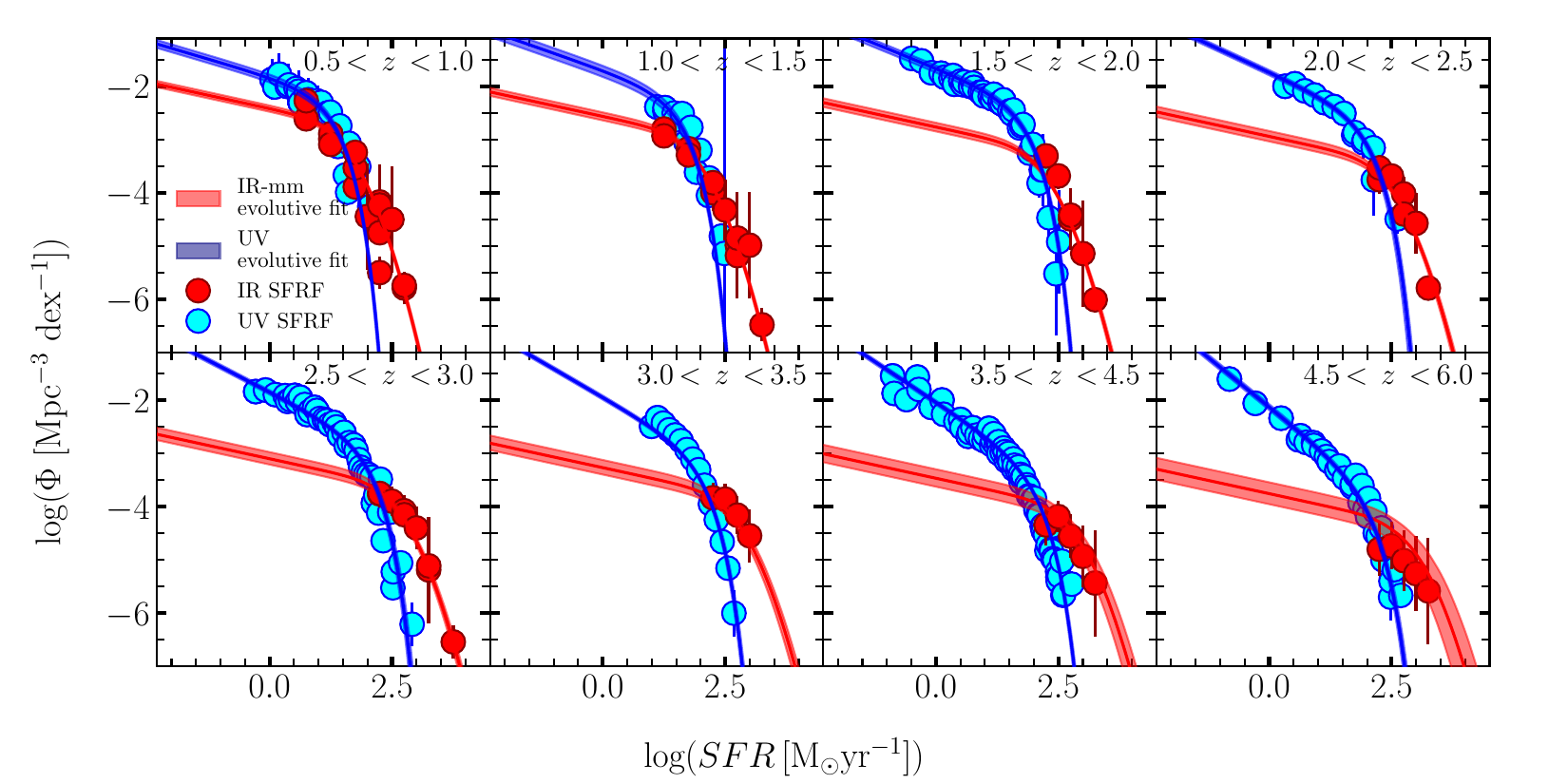}}
{\includegraphics[width=1\textwidth]{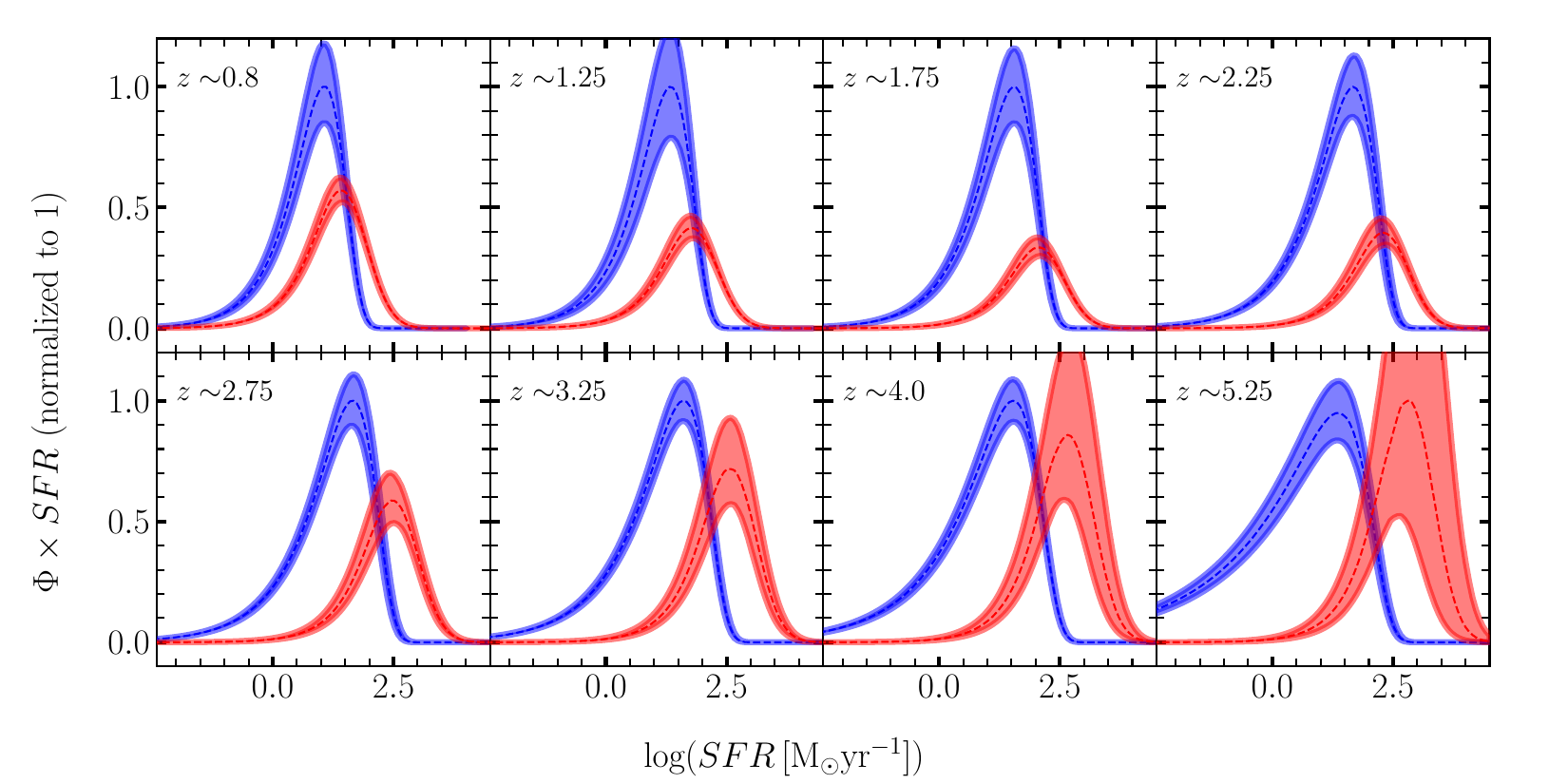}}
\caption{{\it Upper panel:} UV and IR SFRFs at different redshifts. Data points are plotted in cyan (UV) and red (IR), while the best fit is shown as blue and red curves for the UV and IR datasets, respectively. {\it Lower panel:} product between volume density ($\Phi$) and SFR in each redshift bin: this is the quantity that we integrate to obtain the SFRD. The UV (blue area) and the IR (red area) SFR density distributions are compared at different redshifts as function of the SFR. Both curves are normalize to have the peak of the UV at 1.}
\label{fig:SFRF_points_fit}
\end{figure*}

We computed the IR and UV SFRFs at $z \sim 0.5 \,-\,6$, dividing the $z$ range into 8 redshift bins similar to the bins explored by the UV and IR works we are using here. The SFRF UV and IR points are shown in Figure \ref{fig:SFRF_points_fit} (upper panel). Overall, there is a reasonable agreement between UV and IR data points. In particular, at $0 \,<\,z\,<\,1.5$, the points from both tracers are almost overlapped in the range where both have data, with the IR reaching larger values of SFR, while the UV slightly lower ones. At higher redshift, the bright-end of the UV-SFRF starts deviating from that of the IR-SFRF, which decreases less drastically. This discrepancy could be ascribed to the difficulty for UV observations to detect the most dust-obscured galaxies with the highest SFRs (e.g., $SFR \,>\, 10^3$ M$_{\odot}$ yr$^{-1}$). On the other hand, the higher the redshift, the less the IR-mm observations become sensitive enough to detect galaxies at low star forming regimes. In fact, at the lowest redshift bins, UV and IR data points cover more or less the same range in SFR, while, at higher redshift, UV data covers a wider range, especially at the faint-end, but do not reach high SFRs ad the IR data do instead.
\par As a result, the two SFR functions obtained from UV dust-corrected and IR data might display different shapes while yielding similar integrated SFRD. In the next Section, we will discuss these aspects in more detail.

\subsection{SFRF fitting procedure}\label{sec:method}

In order to measure the IR (from the A$^3$COSMOS) and UV (from works in Section \ref{subsec:uv_datasets}) SFRFs, we first fitted our data points (IR and UV separately) by performing a Monte Carlo Markov chain (MCMC) analysis, thus obtaining a best-fit IR and UV SFRF over several redshift bins. We ran the MCMC using the \texttt{PYTHON} package \texttt{emcee} \citep{foremanmackey2013emcee}, which allows us to explore the parameter space at once, using a set of walkers. While the UV SFRFs can be fitted with a classical Schechter function \citep[][]{schechter1976}{}{}, the IR SFRFs typically show an excess in the bright-end, that cannot be simply modeled with a Schechter \citep[][]{lawrence1986irlf,soifer1987irlf,saunders1990modified,rush1993irlf,shupe1998irlf,sanders2003irlf}. For this reason, we adopted a modified Schechter \citep[][]{saunders1990modified}{}{} for the IR-SFRF fit. The classical Schechter is characterized by three free parameters, which are the knee star formation rate (SFR$^*$) and density $\Phi^*$ of the galaxy population, plus the slope of the faint-end ($\alpha$). The modified Schechter has, in addition, a fourth parameter, $\sigma$, describing the shape of the bright-end. The Schechter function is described by the following form:
\begin{equation}
    \Phi(L)dlogL = \Phi^*\left(\frac{L}{L^*}\right)^{\alpha} exp\left[-\frac{L}{L^*}\right]dlogL .
\end{equation}
The modified Schechter can be written as:
\begin{equation}
    \Phi(L)dlogL = \Phi^*\left(\frac{L}{L^*}\right)^{1-\alpha_S} exp\left[-\frac{1}{2\sigma_S^2}log_{10}^2 \left(1+\frac{L}{L^*}\right)\right]dlogL,
\end{equation}

Previous studies \citep[e.g.,][]{caputi2007lfevol,bethermin2011lfevol,gruppioni2013lf}{}{} have shown that these parameters, in particular $L^*$ and $\Phi^*$, evolve with redshift. Indeed, in the IR domain, $L^*$ (proxy of dust-obscured SFR) and $\Phi^*$ of the galaxy population increases and decreases with redshift, respectively \citep[e.g.,][]{gruppioni2013lf,magnelli2013ir}{}{}. This evolution can be described by a functional form, assuming a $z_{\rm break}$ for both parameters, that characterizes the change in their evolution with redshift. We adopted a similar redshift evolution for the UV SFRF, for which we factored in an evolving faint-end slope, following \cite{parsa2016sfrf}. The equations which describe the IR- and UV-SFRF evolutions are reported below:
\begin{equation}
 \begin{cases}
  \Phi^* = \Phi^*_{0} (1+z)^{k_{\rm \rho1}} ~~~~~~~~~~~~~~~~~~~~~~~~ z < z_{\rm \rho0}\\
  \Phi^* = \Phi^*_{0} (1+z)^{k_{\rm \rho2}} (1+z_{\rm \rho0})^{(k_{\rm \rho1}-k_{\rm \rho2})}~z > z_{\rm \rho0}
 \end{cases}
\end{equation}
\begin{equation}
 \begin{cases}
  SFR^* = SFR^*_{0} (1+z)^{k_{\rm L1}} ~~~~~~~~~~~~~~~~~~~~~~~ z < z_{\rm l0}\\
  SFR^* = SFR^*_{0} (1+z)^{k_{\rm L2}} (1+z_{\rm L0})^{(k_{\rm L1}-k_{\rm L2})}~~ z > z_{\rm L0}
 \end{cases}
\end{equation}

where $\Phi^*_{0}$ and SFR$^*_{0}$ are the normalization and characteristic SFR at $z=0$, $k_{\rm \rho1}$, $k_{\rm \rho2}$, $k_{\rm L1}$ and $k_{\rm L2}$ are the exponents for values lower and greater than $z_{\rm \rho0}$ and $z_{\rm L0}$ for $\Phi^*$ and SFR$^*$, respectively. The evolution of $\alpha$ in the UV (dust corrected) SFRF fit, is parametrized by the following power law: 
\begin{equation}
    \alpha (z) = k_{a1} + k_{a2} \times z .
\end{equation}
\par By taking advantage of the wide redshift range and the plethora of independent datasets considered in this analysis, we can constrain the best-fit parameters, identifying the local Schechter function and its evolution. Then, using the functional forms for the evolution described before, we can obtain the best-fit IR and UV-SFRFs for each redshift bin. In this way, we can easily combine SFRF data from different studies at different redshifts (in the same bands and completeness-corrected). In order to accurately constrain the parameters of the best-fit SFRFs, we also combined the A$^3$COSMOS IR-SFRFs with those obtained from Herschel PEP/HerMES on the COSMOS, ECDFS, GOODS-N, and GOODS-S fields \citep[][]{gruppioni2015sam}, containing a much larger number of sources, especially at low redshift, where ALMA does not provide good statistics, as done also by T24 for the IRLFs.
\par The evolutive IR and UV best-fit of the SFRFs are shown in Figure \ref{fig:SFRF_points_fit} (upper panel). Since this fit is not done individually for the data points in each redshift bin, some deviation between the data and model at specific redshift bins might happen; the UV and IR SFRF best-fits are indeed obtained by considering together the data points at all redshifts and finding an evolution compatible with them all. As we mentioned before, there is a discrepancy (from 0.5 to 1 dex in SFR) between the IR SFRF and the UV (dust-corrected) one, confirming that different indicators produce different results \citep[][]{leja2020,katsianis2020,das2024}. Most of this difference is due to the different SFR range traced by the two datasets and to the different slopes at the faint and bright-end of both SFRF best-fits. This issue, is well explained in Figure  \ref{fig:SFRF_points_fit} (lower panel), where we show the SFR density distribution that we integrate to obtain the SFRD (i.e., $\Phi^* \times SFR$). Up to $z \sim 3$, the UV integral is larger than the IR one, while at $z \,>\, 3$ the IR contribution becomes rapidly larger. However, at all redshifts, the two functions cover different intervals of SFR: this means that the same value of the cosmic SFRD could be obtained by two completely different distributions and, consequently, that neither the IR nor the UV alone can account for the total SFRD at any given redshifts. In the next section, we try to address this problem by defining a total IR \& UV SFRF, whose evolution is studied in Section \ref{sfrd}.

\begin{figure*}[]
\centering
{\includegraphics[width=1\textwidth]{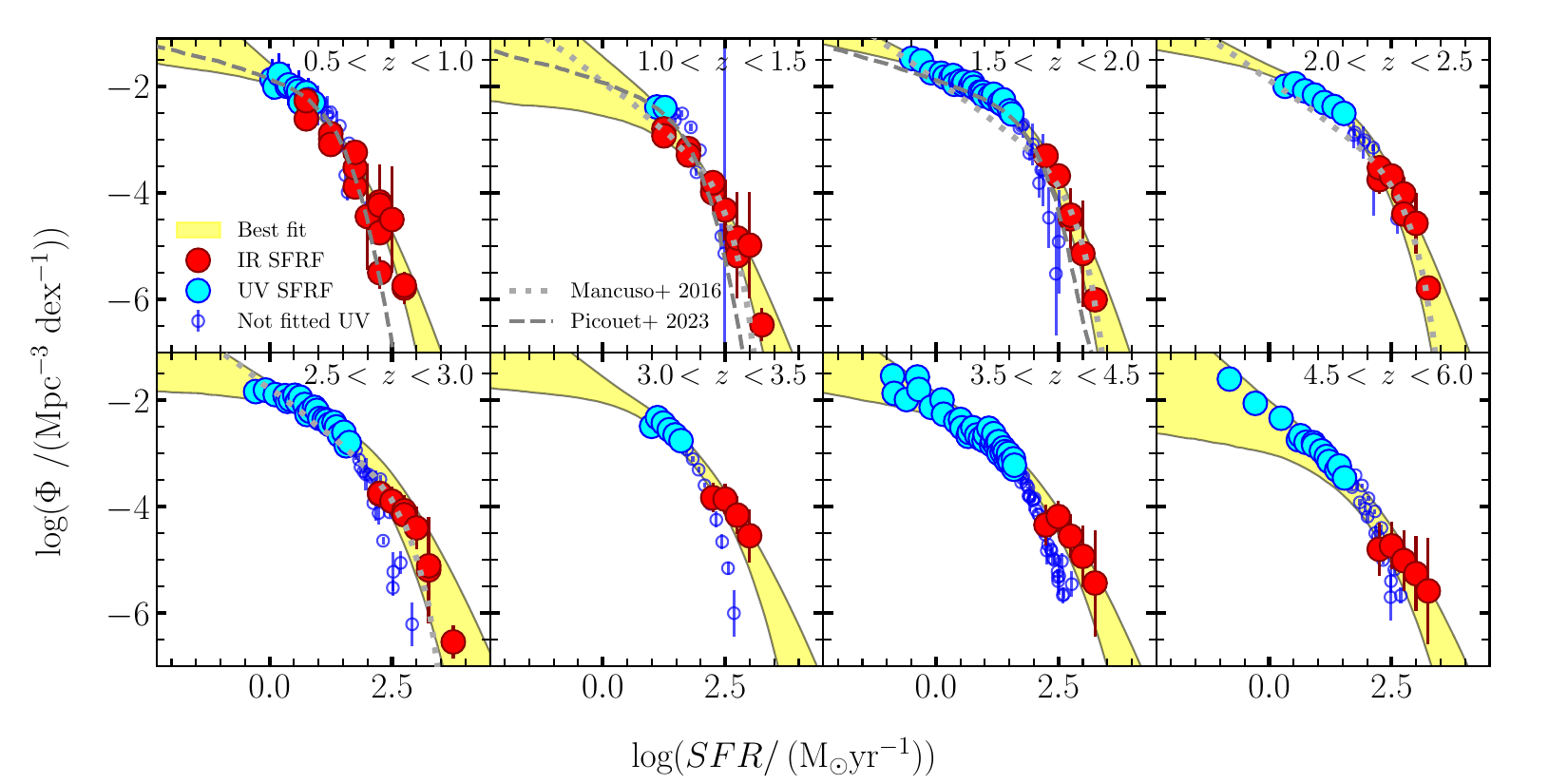}}
\caption{\small{Combined SFRF best-fit. The UV and IR data points are shown as in Figure \ref{fig:SFRF_points_fit}. Blue empty circles are the UV points not used to obtain the combined fit, which is shown as yellow area, corresponding to the 16th and 84 percentiles from the MCMC.}}
\label{fig:SFRF_fit}
\end{figure*}

\begin{figure*}[]
\centering
{\includegraphics[width=1.0\textwidth]{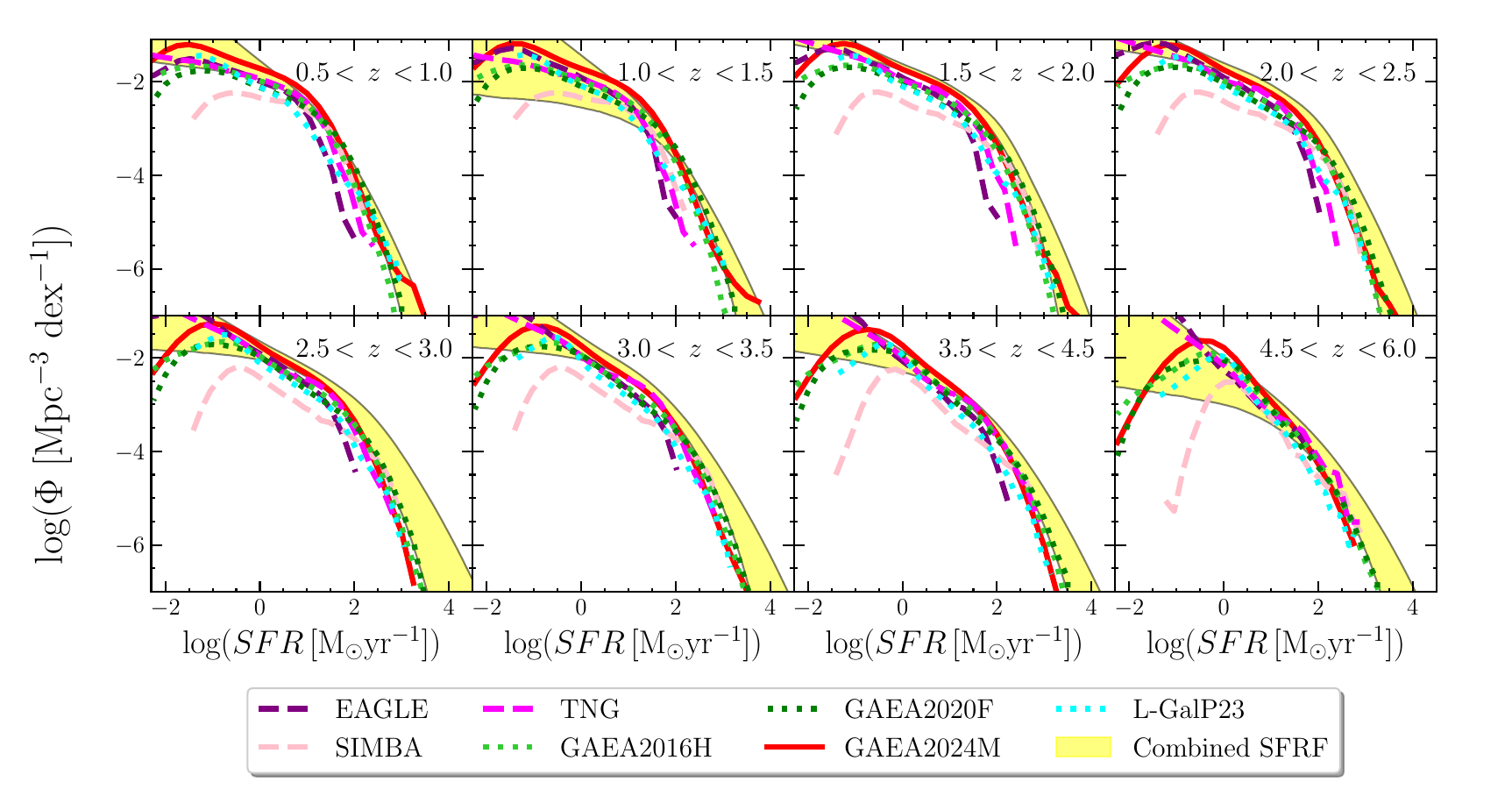}}
\caption{\small{Observed SFRF compared with the prediction from simulations and SAMs. The yellow area represents the observed data used for the fit. The best-fit curve is reported in black. The purple dashed line represents the SFRF from the EAGLE simulation; the pink dashed curve is the result from the SIMBA simulation and the magenta line is from the IllustrisTNG \citep[][]{katsianis2017eagle, katsianis2021simba}. The limegreen and green dotted curves are the prediction by \citet{Hirschmann2016sam} and \citet{fontanot2020sam}, from the GAEA SAM. Finally, the cyan dotted curves are the predictions by \citet{parente2023dust_sam}.}}
\label{fig:LF_sam}
\end{figure*}

\begin{figure}[]
\centering
{\includegraphics[width=0.45\textwidth]{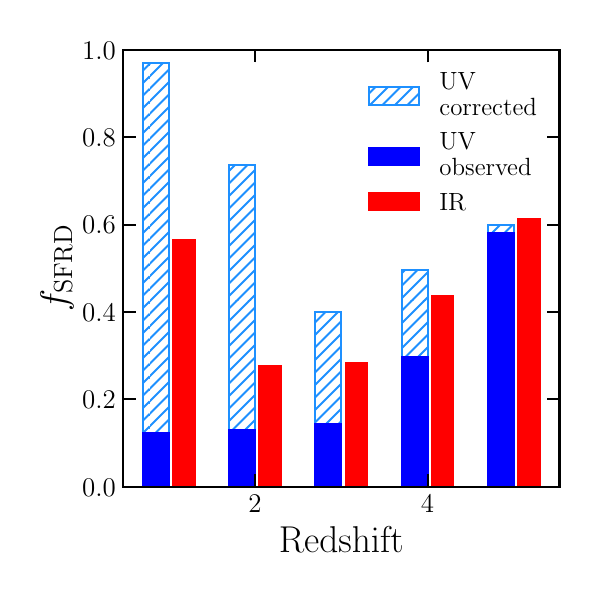}}
\caption{\small{Fraction of the combined SFRD, at each redshift, recovered by UV (light blue and blue histograms) and IR (red histograms) data. The blue filled histograms represent the fraction recovered by the observed UV (i.e., without dust correction), while the dashed light blue histograms correspond to the dust corrected UV). We note that the UV and IR fractions do not sum up to 1, as each SFRD is obtained by integrating the respective best-fit SFR function.}}
\label{fig:sfrd_fraction}
\end{figure}

\begin{figure}[]
\centering
{\includegraphics[width=.5\textwidth]{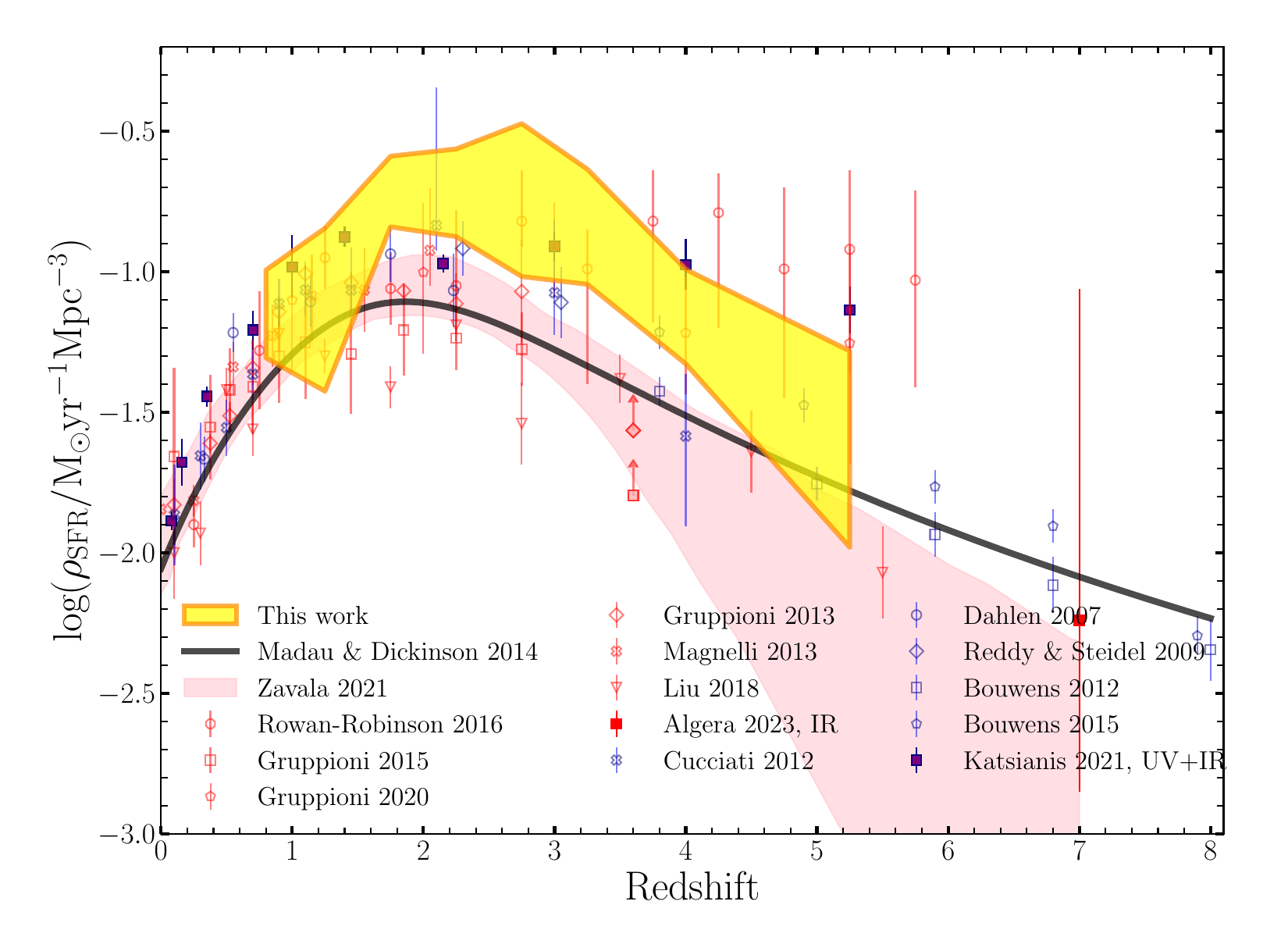}}
{\includegraphics[width=.5\textwidth]{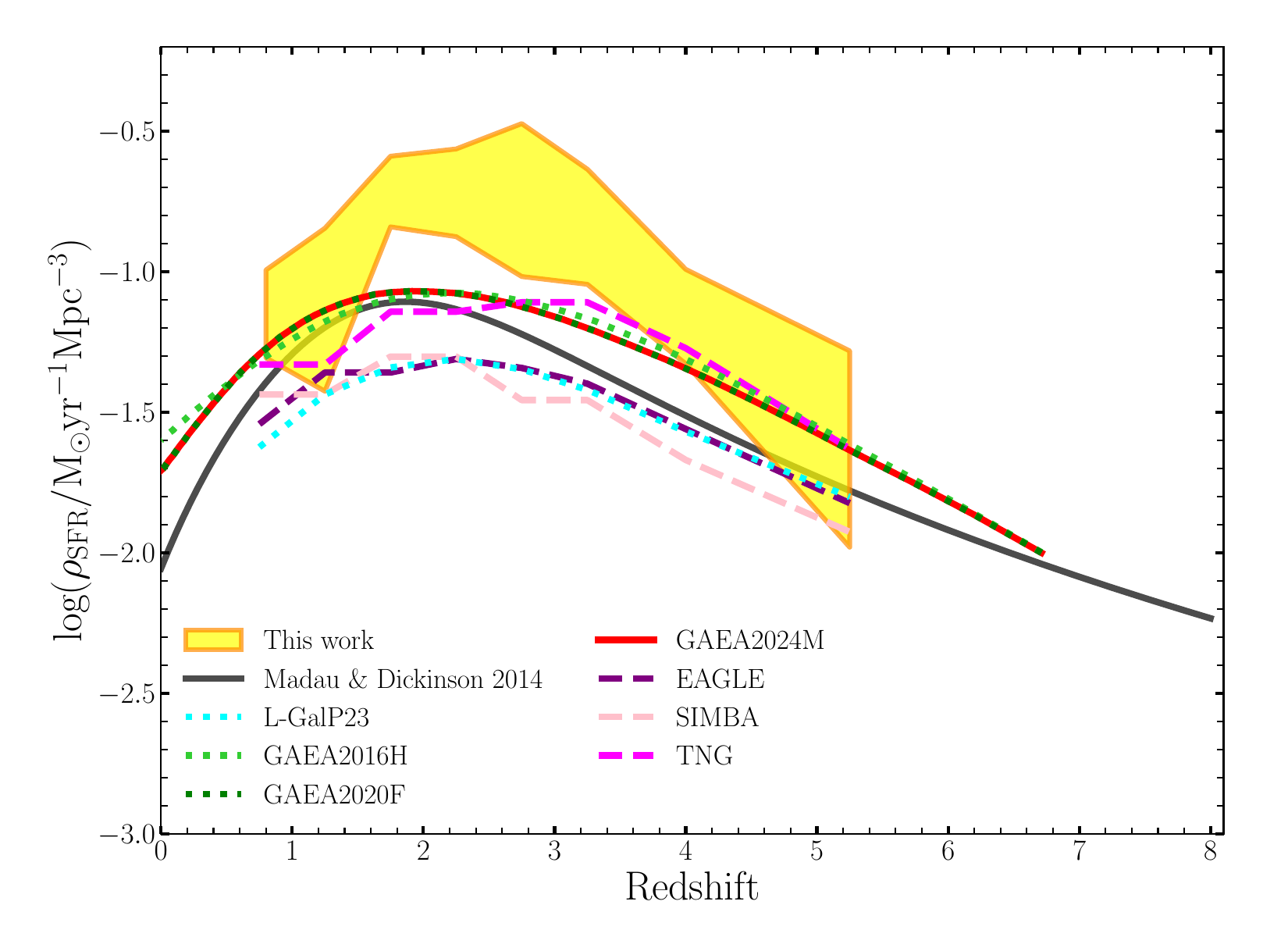}}
\caption{\small{{\it Upper panel:} SFRD obtained by integrating the combined SFRFs at each redshift bin (yellow area), compared to IR and UV results from the literature (plotted as red and blue empty points, respectively). {\it Lower panel:} SFRD compared to predictions by the models reported in this work. The color code of the models is the same as Figure \ref{fig:LF_sam}.}}
\label{fig:sfrd_new}
\end{figure}

\subsection{The IR + UV total SFRF}\label{total_sfrf}
In order to compare our results with those of simulations and SAMs, which predict the total SFRF, we need to trace the same quantity also with data (i.e., a total SFRF). As noticed in the previous section, neither the IR SFRF, or the dust-corrected UV SFRF are able to represent the total SFRF, because, even if corrected for incompleteness and dust attenuation, neither of them samples the whole range of SFRs. We therefore need to derive the best representation possible of the total SFRF, covering all the relevant values of SFRs. To this purpose we have combined the UV and IR SFRF data and performed a MCMC fit under the following assumptions. In principle, to obtain the total SFRF we should use the same galaxy sample and sum the obscured (traced by the IR) and unobscured (traced by the UV not corrected for dust extinction) SFRs for each galaxy. However, as clearly shown by Figure \ref{fig:SFRF_points_fit}, there are no samples of galaxies with observations in the IR and UV covering the exact same interval of SFRs (even though some are partially overlapping). In fact, UV surveys typically sample fainter SFRs, but are not able to detect the most star forming (and obscured) systems, while IR surveys are sensitive to the higher SFRs, but not deep enough to detect the fainter SFR galaxies, containing small amounts of dust. Given the complementarity of the two SFRFs traced by UV and IR, as shown in Figure \ref{fig:SFRF_points_fit}, we assume that the faint-end of the UV-SFRF (corrected for dust attenuation) is tracing $\sim 100\%$ of the SFR at low SFRs and, in a similar way, that the bright-end of the combined SFRF is almost entirely traced by highest SFRs galaxies. With these assumptions, we can therefore fit with a unique function the UV-SFRF data at SFR $<$ SFR$^*_{\rm UV}$ and the observed IR-SFRF at SFRs larger then the IR completeness threshold. We can consider the global fit to data the best representation of the total SFRF. 
\par Before proceeding, in order to check if our assumptions are reasonable and valid in a wide redshift range, we performed a simple test aimed at deriving the combined SFRF by assuming a certain stellar mass function (SMF) and galaxy main sequence (MS) to be compared with our total SFRF from all the data. The details of this test are given in Appendix \ref{app:cutouts} and the result (that confirms the robustness of our method) is shown in Figure \ref{fig:SMF_MS}. 
\par Moreover, a similar approach has already been used in the works by \citet{mancuso2016b,mancuso2016a}, where the authors fit simultaneously the IR and UV (dust-corrected) SFRFs with a Schechter function, using all the IR data points and the UV data points up to $30$ M$_{\odot}$ yr$^{-1}$. In their works, it is clearly shown how the UV and IR traces different regimes of SFR and, therefore, are both needed to compute the total SFRF. Similar findings also emerged from the works by \citet{Bernhard2014, rodriguez_puebla2020ApJ...905..171R}, showing that the IR and UV bands trace very different (almost complementary) regimes of galaxy types in term of stellar masses and SFRs, with the IR tracing typically the most massive, star forming ones and the UV the faintest ones.
\par To perform the fit of the combined UV and IR SFRF we fitted the data in the individual $z$-bin, with a modified Schechter, without fixing the faint and bright-end slopes. In particular, we left $\alpha_S$, $\sigma_S$ (the slope parameters), $\Phi^*$ and $L^*$ free to vary in each bin (see table \ref{tab:fit_res}). In contrast to the previous fits, we choose to fit the data points in each redshift bin individually because of a lack of support from the literature of any hints of evolution for a "total" SFRF. The results of the combined SFRF fit is reported in Figure \ref{fig:SFRF_fit}. In most of the bins, the modified Schechter fits well the data points, both in the UV and IR regimes, enabling us to properly characterize the properties of the SFRF simultaneously in the faint and bright-ends, as well as in the knee. We note, however, that in this parametrization of the SFRF, at $z \,>\, 2$, there is a portion of the SFRF in which we do not have reliable or complete data to fit. Even though with a small impact, this lack of data may influence the final fit because it covers only the range of the SFRF knee. Figure \ref{fig:SFRF_fit} also shows the comparison with the results obtained by \citet{mancuso2016a} and \citet{picouet2023A&A...675A.164P} on the total SFRF, using similar approaches. In particular, in the former work the authors use a compilation of UV and IR LF data to cover both the faint and bright-ends of the SFRF; in the latter, the authors use FIR data from the COSMOS2020 catalog and {\it U}-band observation from the HSC-CLAUDS survey. Our results are consistent with both works, even though comparable up to $z \sim 3$.

\begin{table}[] \label{tab:fit_res}
\renewcommand{\arraystretch}{1.5}
\caption{Best-fit parameters of the total UV+IR SFRFs.}
\begin{tabular}{ccccc}
\hline \hline
\textbf{$z$} & \textbf{$\alpha_S$}    & \textbf{$\sigma_S$}    & \textbf{$\Phi^*$}       & \textbf{$SFR^*$}       \\ \hline
$0.5-1.0$    & $1.71_{-0.13}^{+0.19}$ & $0.71_{-0.08}^{+0.09}$ & $-1.82_{-0.22}^{+0.17}$ & $0.25_{-0.17}^{+0.24}$ \\
$1.0-1.5$    & $1.61_{-0.20}^{+0.31}$ & $0.70_{-0.15}^{+0.17}$ & $-2.38_{-0.49}^{+0.47}$ & $0.71_{-0.46}^{+0.55}$ \\
$1.5-2.0$    & $1.40_{-0.06}^{+0.07}$ & $0.49_{-0.10}^{+0.12}$ & $-2.18_{-0.20}^{+0.22}$ & $1.23_{-0.38}^{+0.34}$ \\
$2.0-2.5$    & $1.38_{-0.08}^{+0.11}$ & $0.56_{-0.13}^{+0.14}$ & $-2.07_{-0.27}^{+0.27}$ & $1.03_{-0.48}^{0.44}$  \\
$2.5-3.0$    & $1.37_{-0.17}^{+0.18}$ & $0.74_{-0.19}^{+0.16}$ & $-2.11_{-0.42}^{+0.23}$ & $0.77_{-0.50}^{+0.73}$ \\
$3.0-3.5$    & $1.41_{-0.20}^{+0.21}$ & $0.77_{-0.13}^{+0.14}$ & $-1.99_{-0.31}^{+0.20}$ & $0.54_{-0.35}^{+0.49}$ \\
$3.5-4.5$    & $1.43_{-0.16}^{+0.17}$ & $0.77_{-0.11}^{+0.11}$ & $-2.22_{-0.29}^{+0.21}$ & $0.48_{-0.32}^{+0.45}$ \\
$4.5-6$      & $1.65_{-0.17}^{+0.27}$ & $0.83_{-0.17}^{+0.16}$ & $-2.79_{-0.55}^{+0.45}$ & $0.68_{-0.45}^{+0.63}$ \\ \hline
\end{tabular}
\tablefoot{Slope parameters ($\alpha_S$ and $\sigma_S$), star formation rate ($SFR^*$) and normalizations ($\Phi^*$) at the knee in the eight redshift bins obtained through the MCMC analysis.}  
\end{table}

\section{Comparison with hydrodynamical simulations and semi-analytical models}\label{sec:comparison_w_models}

We compare our results of the total IR \& UV SFRF with the predictions of state-of-the-art simulations and semi-analytical models. In particular, we discuss the comparison with the SFRF from the EAGLE simulation \citep[][]{schaye2015eagle,crain2015eagle}{}{} by \citet{katsianis2017eagle}, IllustrisTNG \citep[][]{pillepich2018illustrisTNG}{}{} and SIMBA \citep[][]{dave2019simba}{}{} hydrodynamical simulations, the GAEA \citep[][]{Hirschmann2016sam,fontanot2020sam} and \textsc{L-Galaxies} SAMs \citep[][]{henriques2020lgalaxies,parente2023dust_sam}{}{}. In the following, we briefly describe these theoretical frameworks, the main features of each model are summarize in Tables \ref{tab:hydro} and \ref{tab:sam} in Appendix \ref{app:table_models}. The predicted SFRF are shown in Figure \ref{fig:LF_sam}, along with our estimates.

\subsection{Hydrodynamical simulations}

Hydrodynamical simulations aim to follow galaxy formation in a direct and self-consistent manner, by numerically solving the coupled evolution of dark matter and baryonic matter. These simulations incorporate gravitational dynamics, gas hydrodynamics, and sub-grid models for unresolved processes such as star formation and feedback, enabling a detailed characterization of the physical mechanisms that regulate star formation. However, their high computational cost typically limits the accessible cosmological volume, which can impact the statistical sampling of rare objects relevant to the high-SFR end of the SFRF.

\subsubsection{EAGLE}
In the context of stellar formation, the EAGLE simulation incorporates the methodology proposed by \citet{dallavecchia2008sn}, wherein the gaseous medium is categorized into distinct phases: cold molecular clouds, warm atomic gas, and ionized hot gas bubbles. The star formation rate is estimated through the Kennicutt-Schmidt relation \citep[][]{schmidt1959law}{}{}, derived from the surface density of both stars and gas. Supermassive black holes (SMBHs) are included as well, with seeds placed at the center of massive dark matter (DM) halos. The feedback from AGNs is assumed to be responsible for the quenching of massive galaxies and is tuned to reproduce the high mass end of the galaxy stellar mass function (GSMF) at $z = 0$. In this work we used the prediction corresponding to a cosmological box with size $\sim 100$ Mpc. The predictions of the EAGLE simulation are shown in Figure \ref{fig:LF_sam} (magenta dashed curve). The SFR range covered by the predicted SFRF are spanning between $0.01 < {\rm SFR} [{\rm M}_{\odot}{\rm yr}^{-1}] < 2$, thus, we are able to compare it with our results between the faint-end and the knee of the SFRFs. While at lower redshifts the prediction reproduces the faint-end of the combined SFRFs, at $z > 2.5$, it reproduces with better accuracy the knee of the distribution. A good agreement between the observed UV SFRFs and the predictions from EAGLE was found as well by \citet{katsianis2017eagle} at these redshift. In our work we confirm the tension with the higher star formations probed by the IR observations. This discrepancy could be alleviated by decreasing the efficiency of the AGN feedback prescription \citep[][]{katsianis2017eagle,katsianis2017z4}{}{}. It is uncertain though if a larger box-size simulation with better statistics for the high star forming galaxies would reproduce the observations without further re-tuning of the feedback prescriptions. 

\subsubsection{IllustrisTNG}

The IllustrisTNG \citep[][with a box size of $\sim 110$ Mpc]{pillepich2018illustrisTNG}{}{} is an improvement of the cosmological simulation Illustris \citep[][]{genel2014illustris,vogelsberger2014illustris}{}{}, based on the AREPO \citep[][]{springel2010arepo}{}{} code. 
Noteworthy enhancements in the new model include improvements to the prescriptions for supernova and AGN feedback. 
Galactic winds are injected in an isotropic way, with increased wind factors in terms of velocity and energy, resulting in a more efficient overall quenching process in IllustrisTNG compared to its predecessor. Regarding AGN feedback for high BH accretion rates relative to the Eddington limit, the model assumes that a fraction of the accreted rest-mass energy thermally heats the surrounding gas. In cases of low accretion rates, a pure kinetic feedback component is employed, imparting momentum to the surrounding gas in a stochastic manner. We note that unlike Illustris or EAGLE the IllustrisTNG model was {\it constrained} to reproduce the {\it observed} Cosmic star formation rate density (CSFRD) at $z \sim 0-10$ so a good agreement with the SFRF is not guaranteed but somewhat anticipated. The IllustrisTNG SFRF estimates can reach values up to $log({\rm SFR}) \sim 3$, allowing us to compare it with the high star-forming regime of the SFRF. The comparison is similar to that of the EAGLE simulation, with the IllustrisTNG SFRF showing slightly higher values of $\Phi$ at SFR > SFR$^*$. Notably, at $z>3$, where other predictions from cosmological models struggle in reproducing the knee of the combined SFRF, the IllustrisTNG predicts quite well the observed SFRF up to SFR $ \sim 5 \times 10^2$, at $z \sim 5$. We note that similarly to EAGLE, IllustrisTNG suffers significantly from resolution effects and the resulting SFRFs are also impacted by resolution \citep[][]{zhao2020illustrisTNG}{}{}. If the model is run at lower or higher resolution a re-tuning of the feedback prescriptions is required to reproduce the observations (CSFRD and SFRF). In addition, a 100 Mpc box-size is relatively small so it is possible that the simulation is unable to probe rare high star forming objects.

\subsubsection{SIMBA}
The SIMBA \citep[][box size $\sim 150$ Mpc]{dave2019simba}{}{} simulation is the upgraded version of the MUFASA \citep[][]{dave2017mufasa}{}{}, based on the GIZMO code \citep[][]{hopking2015gizmo}{}{}. 
The main enhancement lies in the inclusion of seeding and evolution of BHs, playing a role in the quenching. BH accretion is facilitated through two channels, with a component sourced from cold gas and one originating from hot gas. This propels feedback mechanisms that suppress galaxy activity in the form of kinetic bi-polar outflows and X-ray heating. We note that unlike EAGLE or IllustrisTNG, SIMBA is able to reproduce a bi-modal specific star formation rate function \citep[][]{katsianis2021simba}{}{}. This points that possibly the sophisticated AGN feedback prescription employed is successful at reproducing observables related to galaxy stellar masses and SFRs. As already observed for the previous two simulations, also SIMBA reproduces well the combined SFRF at $z < 2-2.5$, but missing the most star forming population, while having some limitations for the low star forming regime due to its low resolution.

\subsection{Semi-analytical models}

SAMs provide an efficient and flexible framework to explore galaxy formation and evolution. They are built upon dark matter halo merger trees, typically extracted from large N-body simulations, onto which simplified but physically motivated prescriptions for gas cooling, star formation, feedback, and chemical enrichment are implemented. This approach allows SAMs to cover wide cosmological volumes and to rapidly test the impact of different physical assumptions, making them particularly valuable for statistical comparisons with observed galaxy populations such as the SFRF.

\subsubsection{GAEA2016H, GAEA2020F and GAEA2024M}

In this study, we also compare our SFRF results with the predictions from semi-analytic models. Specifically, we employ the predictions from the GAlaxy Evolution and Assembly (GAEA) code, both in its standard realization presented in \citet{Hirschmann2016sam}, GAEA2016H hereafter, and in the updated version from \citet{fontanot2020sam}, GAEA2020F, which correspond to the F06-GAEA run in that paper, based on the Millennium simulation, with a volume $\sim 500^3$ Mpc$^3$.
Both models delineate baryonic evolution within four compartments: the stellar component of galaxies, cold gas within the galactic disk, hot gas in the halo of dark matter, and the ejected gas component. 
This model has been calibrated to reproduce the galaxy stellar mass function up to $z \sim 3$, but it also provide a good agreement with higher redshift observations \citep[][]{fontanot2017highz}{}{}. In more recent rendition of the model \citep[][]{fontanot2020sam}{}{}, the modeling of cold gas accretion onto SMBHs has been improved and better characterized, taking into account several triggering mechanisms (such as mergers and disk instabilities) for the loss of angular momentum of the cold gas, leading to the formation of a reservoir from which its could get accreted to the central BH. Finally, AGN feedback is also introduced in the form of AGN-driven outflows. A new version of the GAEA simulation, the GAEA2024M \citep[][]{delucia2024} includes improvements in the gas accretion onto SMBHs and gas interaction for stripping of satellite galaxies.
\par All the three GAEA2016H, GAEA2020F and GAEA2024M are in good agreement with our data points and best-fit in most of our redshift bins. At higher redshifts, the models begin to diverge from the data, especially at the highest star-forming end (although consistent within the large uncertainties on the fit).

\subsubsection{L-GalP23}
The \textsc{L-Galaxies} SAM \citep[][]{henriques2020lgalaxies}{}{} is the last public release of the Munich galaxy formation model. It is run on top of the Millennium simulation \citep{springel2005DM} and it models the evolution of both the stellar and gaseous components of DM haloes by taking into account a number astrophysical processes, including feedback from SNe and AGNs (in the radio mode).
Here we compare with the predictions of the \textsc{L-Galaxies} model, incorporating updates introduced in \citet{parente2023dust_sam}. These updates include a detailed treatment of dust evolution and a novel prescription for disc instabilities, which considers the instability of both the stellar and gaseous disc, with the latter being capable of inducing starbursts and accreting the central SMBH. Similarly to other models, the L-GalP23 predictions mostly reproduce the faint-end of the SFRF.

The comparison between our results and the simulations described above shows that, despite some difficulties of most models in reproducing the SFRF at the brighter SFRs (at $z > 2$), some progresses have been made in the characterization of the physical processes regulating the star formation in galaxies. Indeed, the agreement between data and models has significantly improved with respect to similar works carried out in the past years \citep[e.g.,][]{gruppioni2015sam}{}{}, now allowing a good agreement with data up to $z \sim 3.5-4$. However, we note that all the models considered predict lower SFRF values at SFR $>10^3$ M$_{\sun}$yr$^{-1}$. As noted by \citep[][]{katsianis2021sfrd}{}{}, it is always possible that our IR SFRFs at these regimes are over-estimated due to: {\it i)} overestimation due to dust being heated by old populations not relevant to current star-formation {\it ii)} overestimation due to larger polycyclic aromatic hydrocarbons emission of distant galaxies {\it iii)} ultra-luminous IR galaxies at high redshift being offset from the typically used SFR calibration adopted by \citet{kennicutt1998sfr} {\it iv)} no consideration of the Eddington Bias at the high star forming end {\it v)} residuals of AGN activity in the most IR luminous sources. 
However, these effects should account for an almost negligible fraction of the IR luminosity. In particular, effects such as the Eddington bias \citep[which may affects the bright-end of the SFRF, see e.g.,][]{ilbert2013cosmos,picouet2023A&A...675A.164P} may represents a subdominant source of uncertainties \citep[$\sim 0.1$dex, see e.g.,][]{caputi2011MNRAS.413..162C}, especially in our case, where the uncertainties on the SFRF fit are largely dominated by the errors on the photometric redshifts of the sample and the errors on the IR luminosities, which have been taken into account by performing a bootstrap resampling of the SFRs within their errors. An additional source of uncertainty that may lead to some level of discrepancy between observed data and simulation or models is the effect of metallicity on the conversions typically used in the literature (and also in this work) to obtain the SFR from a luminosity. Indeed, it has been shown by, for instances, \cite{madau2014sfrd, tacchella2018ApJ...868...92T,rodriguez_puebla2020ApJ...905..171R}, that a not proper account for the metallicity can lead to errors up to a factor $\sim0.5$ dex. In our case however, for comparison purposes, we nonetheless used standard calibration factors. Finally, the effect of cosmic variance, as a source of bias in this analysis, has been shown to be almost negligible in surveys composed by several pointings \citep[][]{driver?robotham2010MNRAS.407.2131D}. In particular \citet{adscheid2024a3cosmos} shown that its effect on the \a3 can be considered negligible.

\section{The total SFRD}\label{sfrd}
In this Section, we present the cosmic SFRD obtained by integrating the combined SFRFs and compare the SFRD$_{\rm TOT}$ to the estimates we obtain when integrating the predictions by simulations and SAMs.

\subsection{Total SFRD}
To compute the total SFRD, we integrated the total SFRF down to $SFR = 0.03 \times SFR^*$, following the approach by \citet{madau2014sfrd}. We investigated the fraction of the total SFRD that can be estimated by either the IR or the UV (dust-corrected) SFRDs. This comparison is shown in Figure \ref{fig:sfrd_fraction}. We stress that, in this comparison, the UV and IR SFRDs do not sum up to 1, because the figure shows what fraction of the total SFRD is retrieved by using either the IR or the UV probes. The UV dust-corrected SFRD, obtained by integrating the UV SFRF, is able to recover most of the total SFRD at $z \sim 1$, but the ratio between the UV and total SFRD decreases at higher redshifts, becoming compatible with the IR estimates. The IR SFRD is instead able to probe about $50 \%$ of the SFRD at $z \sim 1$, but its contribution increases up to $\sim 60 \%$ at $z \sim 5$. The dust-corrections are crucial at lower redshift, where the SFRD from the observed UV is a very small fraction, but become less important going towards higher redhsifts, where the observed UV is compatible with the IR SFRD. Thus, the combination of UV and IR is fundamental to retrieve the $100 \%$ of the SFRD at $ z \gtrsim 1$.
\par The total SFRD obtained by integrating our total SFRF is presented in Figure \ref{fig:sfrd_new}. In the upper panel, we compare it with results from UV and IR works in the literature. Up to $ z \sim 1.5$, where the UV and IR SFRFs are still similar, our result is in agreement with most of the UV and IR estimates. From $z \sim 2$ to $z \sim 4$, the total SFRD is $\sim 2$ times higher than previous estimates at the same redshifts from either IR or UV only, while at $z \gtrsim 4.5$ is broadly consistent with some of the UV and IR-only derivations.

\subsection{Comparison with models}
Similarly to the SFRF, we compared the total SFRD with the SFRD predictions from the models considered in this work. As expected, the EAGLE SFRD (purple line) is not able to reproduce the observed SFRD in the covered SFR range as it does not extend to values higher than $\sim 100$ M$_{\odot}$ yr$^{-1}$. This limitation was already highlighted in \citet{furlong2015eagle}. The SFRD from the SIMBA simulation (pink curve) shows a similar trend as for the EAGLE, being lower than the combined SFRD. The IllustrisTNG (magenta dashed line, whose subgrid parameters were tuned to reproduce the observed CSFRD) simulation is instead able to reproduce the SFRD at high redshift ($z>3$), while the GAEA SFRDs are consistent with our results up to $z \sim 1.5$ and lower at higher redshifts, predicting a SFRD higher than that derived by \citet{madau2014sfrd}. At $z \sim 4-5$ GAEA is again consistent with our result. Finally, the modified version of the \texttt{L-GalP23} models by \citet{parente2023dust_sam} is producing a SFRD similar to those obtained using the EAGLE and SIMBA simulations. In addition, at low redshifts ($z < 1$), semi-analytical models appear to predict a higher SFRD than observed. This discrepancy may be attributed to an overproduction of star-forming galaxies in the local universe by the SAMs \citep[see e.g.,][]{collacchioni2018}. Conversely, at high redshifts ($z > 2.5$), there seems to be a deficiency in the fraction of galaxies exhibiting particularly active star formation (e.g., SFR $> 1000$ M$_{\odot}$ yr$^{-1}$), characterizing the observed bright end of the IR-SFRF at those redshifts. We note that \citet{gruppioni2015sam} showed that a list of SAMs \citep[e.g.,][]{monaco2007sam,henriques2015sam,delucia2007sam}{}{} was able to reproduce the bright end only up to $z<1-1.5$, while the recent models by \citet{Hirschmann2016sam}, \citet{fontanot2020sam} and \citet{delucia2024} are now consistent with the observations up to $z \sim 2-2.5$. This highlights the improvement of SAMs in the last ten years.

\begin{figure}[]
\centering
{\includegraphics[width=.5\textwidth]{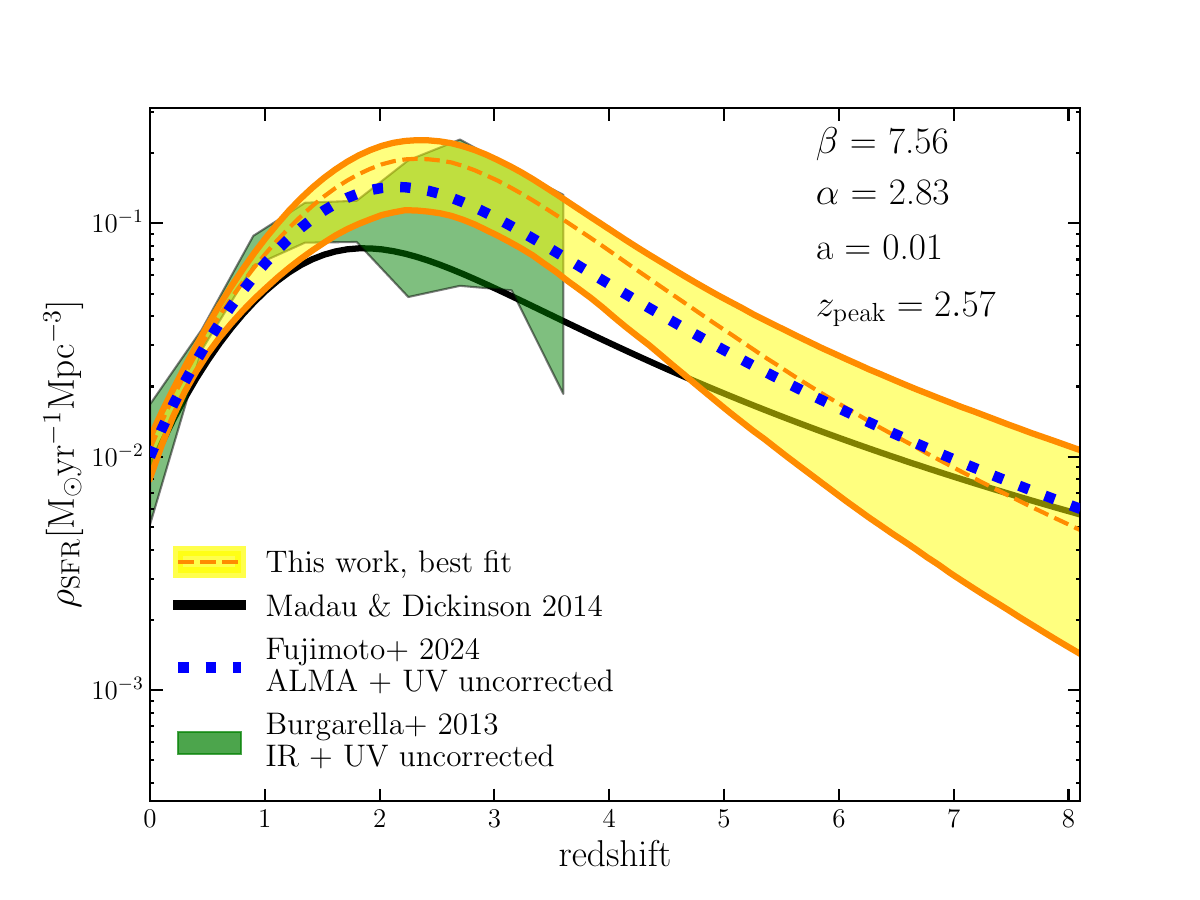}}
\caption{\small{Fit to the total SFRD using the functional form by \citet{madau2014sfrd}. The best fit to our data is the orange dashed line, with yellow errorbands. The fit by \citet{madau2014sfrd} is reported in black. A similar fit by \citet{fujimoto2024sfrd} is represented by the blue dotted line, while the green shaded area is the combined SFRD estimated by \citet{burgarella2013sfrd}. In the upper-right corner, we display the best-fit values of the free parameters of Equation \ref{eq:sfrd_fit}.}}
\label{fig:sfrd_fitted}
\end{figure}

\subsection{A fit of the cosmic SFRD}
In this section, we provide a new fit to the observed total SFRD, adopting the same parametrization described in \citet{madau2014sfrd} \citep[even though some other parameterizations, with Gamma functions or log-normal form, have been shown to be more physically motivated, see e.g.,][]{gladders2013,katsianis2023}. In their work, \citet{madau2014sfrd} fit the observed UV and IR SFRD with the following functional form:
\begin{equation}\label{eq:sfrd_fit}
    \psi(z) = a  \frac{(1+z)^\alpha}{1+[(1+z)/(1+z_{\rm peak})]^{\beta}} \,\, \text{M$_{\odot}$ yr$^{-1}$ Mpc$^{-3}$,}
\end{equation}
where $a$ is the normalization at $z = 0$, $z_{\rm peak}$ is the redshift of the peak, $\alpha$ is the slope of the SFRD at $z < z_{\rm peak}$ and $\beta$ regulates the slope at $z > z_{\rm peak}$. We fitted the total IR \& UV SFRD with an MCMC fitting procedure, leaving the four parameters free to vary. The result is shown in Figure \ref{fig:sfrd_fitted}. We obtain a slightly lower normalization with respect to what is found by \citet{madau2014sfrd} ($a = 0.01_{-0.001}^{+0.001}$, $a_{\rm MD14} = 0.015$); the slopes are also quite different, with $\alpha = 2.83_{-0.035}^{+0.037}$ ($\alpha_{\rm MD14} = 2.7$) and $\beta = 7.56_{-1.10}^{+0.37}$ ($\beta_{\rm MD14} = 5.6$). The main difference is a more pronounced peak, falling at higher redshift ($z_{\rm peak} = 2.57_{-0.38}^{+0.34}$, with respect to $z_{\rm peak,MD14} = 1.9$). This may be due to more accurate IR measurements available today, allowing to probe dust-obscured star formation up to highr redshifts \citep[e.g.,][]{algera2023,liu2025,sun2025}. 
\par Our results are in agreement with the works from \citet{burgarella2013sfrd, fujimoto2024sfrd} which use an IR + UV comprehensive approach. The first work is carried out by deriving the LFs from the {\it Herschel}/PEP+HerMES surveys (FIR) \citep[][]{gruppioni2013lf} and for the UV using data from the VVDS \citep[][]{cucciati2012sfrd}. The authors compute the combined SFRD by summing up the dust-obscured contribution (derived from the FIR) and the unobscured contribution (from the FUV), at $0 \,<\,z\,<\,3.5$ (covering the peak of the SFRD, see Figure \ref{fig:sfrd_fitted}). They find the SFRD to rise between $z \sim 0$ and $z \sim 1.5$, then to softly decrease up to $z \sim 3.5$. The total SFRD derived by \citet{burgarella2013sfrd} has an higher normalization at the peak with respect to \citet{madau2014sfrd}, in agreement with our findings of an higher SFRD at $2 \,<\,z\,<\,3$. The second work, by \citet{fujimoto2024sfrd}, is obtained from the ALMA Lensing Cluster Survey, covering a broad redshift range ($1 \,<\,z\,<\,8$). Unlike the work by \citet{burgarella2013sfrd}, they do not combine IR with UV LFs, but, thanks to the excellent sensitivity reached by ALMA in a lensed field, they were able to probe lower values of the IR-LF also at very high redshifts. In this way, by sampling the IR-LF down to its faint-end, they were able to properly estimate $\alpha$, finding a steeper value than what is typically fitted to the IR-LF \citep[see e.g.,][]{gruppioni2013lf, gruppioni2020alpine, traina2024sfrd}.  By combining their result on the dust-obscured SFRD, with the unobscured SFRD from \cite{bouwens2020sfrd}, they obtain the total SFRD over a very broad redshift range. In Figure \ref{fig:sfrd_fitted}, we show the fit to their data, performed with the same functional form as our fit. They find a normalization ($a = 0.01$) and redshift peak of the SFRD ($z_{\rm peak} \sim 2.3$ in a very good agreement with our results. We note that \citet{fujimoto2024sfrd} report the estimates of the SFRD using two different lower limits for the integration of the IR-LF. In particular, the fit in Figure \ref{fig:sfrd_fitted} is obtained by integrating the IR-LF down to $L_{\rm IR} = 10^{10}$ L$_{\odot}$, which is slightly higher than our lower integration limits (spanning from $\sim 4\times10^8$ to $\sim 4\times10^9$ L$_{\odot}$). However, the authors also consider a less conservative integration limit, $L_{\rm IR} = 10^{8}$ L$_{\odot}$, which leads to higher estimate for the total SFRD ($\sim 3 \times$ larger SFRD, see Figure 18 of their paper), even more consistent with our results.

\section{Summary and Conclusions}\label{endings}

By combining the recent IR-LF and SFRD obtained by \citet{traina2024sfrd} with the dust corrected UV-SFRF derived in various literature works, we estimated the total (IR and UV) SFRFs and SFRD over the $z = 0.5 - 6$ redshift range. Furthermore, we compared these results with predictions from state-of-the-art SAMs and hydrodinamical simulations to assess how well these models can reproduce the observational estimates. The obtained results can be summarized as follows:

\begin{itemize}

   \item The IR and UV dust-corrected star formation functions sample different ranges of SFRs. Indeed, the UV-SFRF extends to values below 10 M$_{\odot}$yr$^{-1}$ up to $z \sim 2$ and reaches values of 100 M$_{\odot}$yr$^{-1}$ at higher redshifts. On the other hand, the IR-SFRF, dominates at higher SFRs, particularly in the range of 100-1000 M$_{\odot}$yr$^{-1}$ \citep[in agreement with previous works, e.g.,][]{katsianis2017z4,katsianis2021sfrd}.
   
   \item Using the dust-corrected UV-SFRF data points at values of SFR lower than the knee and the IR-SFRF at values higher than the knee, we derived the total best-fitting SFRF. The comparison the total SFRF with predictions from simulations and SAMs reveals that the latter are capable of reproducing the faint-end quite well at all redshifts, but the bright-end only up to $z \sim 2-2.5$. At higher redshifts in fact the models show a lack of galaxies with very high SFR, responsible for the bright-end of the observed SFRF.

  \item The total SFRD shows a typical increasing trend at $0.5<z<1.5$, a broad peak up to $z \sim 3$ and a decrease towards $z \sim 6$. The models predict an SFRD consistent with observed values up to approximately $z \sim 2$. At higher redshifts, they are consistent within the errors. The ratios between IR and total SFRD and UV (dust-corrected) and total SFRD are useful to quantify how good are both tracers in reproducing the evolution of the total SFRD. We found that both of them are able to reproduce it up to $z \sim 1$, but, at higher redshifts, they only accounts for $\sim 60\%$ of the total SFRD.

  \item We fitted the total SFRD with the same functional form adopted by \citet{madau2014sfrd} and \citet{fujimoto2024sfrd}. We find the peak of the SFRD to be located at higher redshifts ($z_{\rm peak} \sim 2.6$) and to be higher in the normalization of the peak, with respect to what is obtained by \citet{madau2014sfrd}. Our results are instead consistent with the observed total SFRD by \citet{burgarella2013sfrd} and the recent results by \citet{fujimoto2024sfrd}, which find a normalization and redshift of the peak very similar to ours.
  
\end{itemize}

\begin{acknowledgements} 
AT acknowledges support from the “INAF Ricerca Fondamentale 2023 – Large GO” grant. AK is supported by the Guangdong Basic and applied Basic research foundation (No: 2025A1515012670) and the 100 talents program of Sun Yat-sen University. SA gratefully acknowledges the Collaborative Research Center 1601 (SFB 1601 sub-project C2) funded by the Deutsche Forschungsgemeinschaft (DFG, German Research Foundation) – 500700252. ID acknowledges funding by the European Union – NextGenerationEU, RRF M4C2 1.1, Project 2022JZJBHM: "AGN-sCAN: zooming-in on the AGN-galaxy connection since the cosmic noon" - CUP C53D23001120006.
\end{acknowledgements}

\bibliographystyle{aa}
\bibliography{1biblio}

@ARTICLE{bouwens2014sfrd,
       author = {{Bouwens}, R.~J. and {Bradley}, L. and {Zitrin}, A. and {Coe}, D. and {Franx}, M. and {Zheng}, W. and {Smit}, R. and {Host}, O. and {Postman}, M. and {Moustakas}, L. and {Labb{\'e}}, I. and {Carrasco}, M. and {Molino}, A. and {Donahue}, M. and {Kelson}, D.~D. and {Meneghetti}, M. and {Ben{\'\i}tez}, N. and {Lemze}, D. and {Umetsu}, K. and {Broadhurst}, T. and {Moustakas}, J. and {Rosati}, P. and {Jouvel}, S. and {Bartelmann}, M. and {Ford}, H. and {Graves}, G. and {Grillo}, C. and {Infante}, L. and {Jimenez-Teja}, Y. and {Lahav}, O. and {Maoz}, D. and {Medezinski}, E. and {Melchior}, P. and {Merten}, J. and {Nonino}, M. and {Ogaz}, S. and {Seitz}, S.},
        title = "{A Census of Star-forming Galaxies in the Z \raisebox{-0.5ex}\textasciitilde 9-10 Universe based on HST+Spitzer Observations over 19 Clash Clusters: Three Candidate Z \raisebox{-0.5ex}\textasciitilde 9-10 Galaxies and Improved Constraints on the Star Formation Rate Density at Z \raisebox{-0.5ex}\textasciitilde 9.2}",
      journal = {\apj},
         year = 2014,
        month = nov,
       volume = {795},
       number = {2},
          eid = {126},
        pages = {126},
          doi = {10.1088/0004-637X/795/2/126},
archivePrefix = {arXiv},
       eprint = {1211.2230},
 primaryClass = {astro-ph.CO},
       adsurl = {https://ui.adsabs.harvard.edu/abs/2014ApJ...795..126B}
}

@ARTICLE{oesch2015sfrd,
       author = {{Oesch}, P.~A. and {Bouwens}, R.~J. and {Illingworth}, G.~D. and {Franx}, M. and {Ammons}, S.~M. and {van Dokkum}, P.~G. and {Trenti}, M. and {Labb{\'e}}, I.},
        title = "{First Frontier Field Constraints on the Cosmic Star Formation Rate Density at z {\ensuremath{\sim}} 10{\textemdash}The Impact of Lensing Shear on Completeness of High-redshift Galaxy Samples}",
      journal = {\apj},
         year = 2015,
        month = jul,
       volume = {808},
       number = {1},
          eid = {104},
        pages = {104},
          doi = {10.1088/0004-637X/808/1/104},
archivePrefix = {arXiv},
       eprint = {1409.1228},
 primaryClass = {astro-ph.GA},
       adsurl = {https://ui.adsabs.harvard.edu/abs/2015ApJ...808..104O}
}

@ARTICLE{oesch2018sfrd,
       author = {{Oesch}, P.~A. and {Bouwens}, R.~J. and {Illingworth}, G.~D. and {Labb{\'e}}, I. and {Stefanon}, M.},
        title = "{The Dearth of z {\ensuremath{\sim}} 10 Galaxies in All HST Legacy Fields{\textemdash}The Rapid Evolution of the Galaxy Population in the First 500 Myr}",
      journal = {\apj},
         year = 2018,
        month = mar,
       volume = {855},
       number = {2},
          eid = {105},
        pages = {105},
          doi = {10.3847/1538-4357/aab03f},
archivePrefix = {arXiv},
       eprint = {1710.11131},
 primaryClass = {astro-ph.GA},
       adsurl = {https://ui.adsabs.harvard.edu/abs/2018ApJ...855..105O}
}

@ARTICLE{laporte2016sfrd,
       author = {{Laporte}, N. and {Infante}, L. and {Troncoso Iribarren}, P. and {Zheng}, W. and {Molino}, A. and {Bauer}, F.~E. and {Bina}, D. and {Broadhurst}, Tom and {Chilingarian}, I. and {Huang}, X. and {Garcia}, S. and {Kim}, S. and {Marques-Chaves}, R. and {Moustakas}, J. and {Pell{\'o}}, R. and {P{\'e}rez-Fournon}, I. and {Shu}, X. and {Streblyanska}, A. and {Zitrin}, A.},
        title = "{Young Galaxy Candidates in the Hubble Frontier Fields. III. MACS J0717.5+3745}",
      journal = {\apj},
         year = 2016,
        month = apr,
       volume = {820},
       number = {2},
          eid = {98},
        pages = {98},
          doi = {10.3847/0004-637X/820/2/98},
archivePrefix = {arXiv},
       eprint = {1602.02775},
 primaryClass = {astro-ph.GA},
       adsurl = {https://ui.adsabs.harvard.edu/abs/2016ApJ...820...98L}
}

@ARTICLE{magnelli2013ir,
       author = {{Magnelli}, B. and {Popesso}, P. and {Berta}, S. and {Pozzi}, F. and {Elbaz}, D. and {Lutz}, D. and {Dickinson}, M. and {Altieri}, B. and {Andreani}, P. and {Aussel}, H. and {B{\'e}thermin}, M. and {Bongiovanni}, A. and {Cepa}, J. and {Charmandaris}, V. and {Chary}, R. -R. and {Cimatti}, A. and {Daddi}, E. and {F{\"o}rster Schreiber}, N.~M. and {Genzel}, R. and {Gruppioni}, C. and {Harwit}, M. and {Hwang}, H.~S. and {Ivison}, R.~J. and {Magdis}, G. and {Maiolino}, R. and {Murphy}, E. and {Nordon}, R. and {Pannella}, M. and {P{\'e}rez Garc{\'\i}a}, A. and {Poglitsch}, A. and {Rosario}, D. and {Sanchez-Portal}, M. and {Santini}, P. and {Scott}, D. and {Sturm}, E. and {Tacconi}, L.~J. and {Valtchanov}, I.},
        title = "{The deepest Herschel-PACS far-infrared survey: number counts and infrared luminosity functions from combined PEP/GOODS-H observations}",
      journal = {\aap},
         year = 2013,
        month = may,
       volume = {553},
          eid = {A132},
        pages = {A132},
          doi = {10.1051/0004-6361/201321371},
archivePrefix = {arXiv},
       eprint = {1303.4436},
 primaryClass = {astro-ph.CO},
       adsurl = {https://ui.adsabs.harvard.edu/abs/2013A&A...553A.132M}
}

@ARTICLE{gruppioni2020alpine,
       author = {{Gruppioni}, C. and {B{\'e}thermin}, M. and {Loiacono}, F. and {Le F{\`e}vre}, O. and {Capak}, P. and {Cassata}, P. and {Faisst}, A.~L. and {Schaerer}, D. and {Silverman}, J. and {Yan}, L. and {Bardelli}, S. and {Boquien}, M. and {Carraro}, R. and {Cimatti}, A. and {Dessauges-Zavadsky}, M. and {Ginolfi}, M. and {Fujimoto}, S. and {Hathi}, N.~P. and {Jones}, G.~C. and {Khusanova}, Y. and {Koekemoer}, A.~M. and {Lagache}, G. and {Lemaux}, B.~C. and {Oesch}, P.~A. and {Pozzi}, F. and {Riechers}, D.~A. and {Rodighiero}, G. and {Romano}, M. and {Talia}, M. and {Vallini}, L. and {Vergani}, D. and {Zamorani}, G. and {Zucca}, E.},
        title = "{The ALPINE-ALMA [CII] survey. The nature, luminosity function, and star formation history of dusty galaxies up to z ≃ 6}",
      journal = {\aap},
         year = 2020,
        month = nov,
       volume = {643},
          eid = {A8},
        pages = {A8},
          doi = {10.1051/0004-6361/202038487},
archivePrefix = {arXiv},
       eprint = {2006.04974},
 primaryClass = {astro-ph.GA},
       adsurl = {https://ui.adsabs.harvard.edu/abs/2020A&A...643A...8G}
}

@ARTICLE{madau2014sfrd,
       author = {{Madau}, Piero and {Dickinson}, Mark},
        title = "{Cosmic Star-Formation History}",
      journal = {\araa},
         year = 2014,
        month = aug,
       volume = {52},
        pages = {415-486},
          doi = {10.1146/annurev-astro-081811-125615},
archivePrefix = {arXiv},
       eprint = {1403.0007},
 primaryClass = {astro-ph.CO},
       adsurl = {https://ui.adsabs.harvard.edu/abs/2014ARA&A..52..415M}
}

@ARTICLE{boquien2019cigale,
       author = {{Boquien}, M. and {Burgarella}, D. and {Roehlly}, Y. and {Buat}, V. and {Ciesla}, L. and {Corre}, D. and {Inoue}, A.~K. and {Salas}, H.},
        title = "{CIGALE: a python Code Investigating GALaxy Emission}",
      journal = {\aap},
         year = 2019,
        month = feb,
       volume = {622},
          eid = {A103},
        pages = {A103},
          doi = {10.1051/0004-6361/201834156},
archivePrefix = {arXiv},
       eprint = {1811.03094},
 primaryClass = {astro-ph.GA},
       adsurl = {https://ui.adsabs.harvard.edu/abs/2019A&A...622A.103B}
}

@ARTICLE{weaver2022cosmos2020,
       author = {{Weaver}, J.~R. and {Kauffmann}, O.~B. and {Ilbert}, O. and {McCracken}, H.~J. and {Moneti}, A. and {Toft}, S. and {Brammer}, G. and {Shuntov}, M. and {Davidzon}, I. and {Hsieh}, B.~C. and {Laigle}, C. and {Anastasiou}, A. and {Jespersen}, C.~K. and {Vinther}, J. and {Capak}, P. and {Casey}, C.~M. and {McPartland}, C.~J.~R. and {Milvang-Jensen}, B. and {Mobasher}, B. and {Sanders}, D.~B. and {Zalesky}, L. and {Arnouts}, S. and {Aussel}, H. and {Dunlop}, J.~S. and {Faisst}, A. and {Franx}, M. and {Furtak}, L.~J. and {Fynbo}, J.~P.~U. and {Gould}, K.~M.~L. and {Greve}, T.~R. and {Gwyn}, S. and {Kartaltepe}, J.~S. and {Kashino}, D. and {Koekemoer}, A.~M. and {Kokorev}, V. and {Le F{\`e}vre}, O. and {Lilly}, S. and {Masters}, D. and {Magdis}, G. and {Mehta}, V. and {Peng}, Y. and {Riechers}, D.~A. and {Salvato}, M. and {Sawicki}, M. and {Scarlata}, C. and {Scoville}, N. and {Shirley}, R. and {Silverman}, J.~D. and {Sneppen}, A. and {Smolc̆i{\'c}}, V. and {Steinhardt}, C. and {Stern}, D. and {Tanaka}, M. and {Taniguchi}, Y. and {Teplitz}, H.~I. and {Vaccari}, M. and {Wang}, W. -H. and {Zamorani}, G.},
        title = "{COSMOS2020: A Panchromatic View of the Universe to z 10 from Two Complementary Catalogs}",
      journal = {\apjs},
         year = 2022,
        month = jan,
       volume = {258},
       number = {1},
          eid = {11},
        pages = {11},
          doi = {10.3847/1538-4365/ac3078},
archivePrefix = {arXiv},
       eprint = {2110.13923},
 primaryClass = {astro-ph.GA},
       adsurl = {https://ui.adsabs.harvard.edu/abs/2022ApJS..258...11W}
}

@ARTICLE{liu2019a31,
       author = {{Liu}, Daizhong and {Schinnerer}, E. and {Groves}, B. and {Magnelli}, B. and {Lang}, P. and {Leslie}, S. and {Jim{\'e}nez-Andrade}, E. and {Riechers}, D.~A. and {Popping}, G. and {Magdis}, Georgios E. and {Daddi}, E. and {Sargent}, M. and {Gao}, Yu and {Fudamoto}, Y. and {Oesch}, P.~A. and {Bertoldi}, F.},
        title = "{Automated Mining of the ALMA Archive in the COSMOS Field (A$^{3}$COSMOS). II. Cold Molecular Gas Evolution out to Redshift 6}",
      journal = {\apj},
         year = 2019,
        month = dec,
       volume = {887},
       number = {2},
          eid = {235},
        pages = {235},
          doi = {10.3847/1538-4357/ab578d},
archivePrefix = {arXiv},
       eprint = {1910.12883},
 primaryClass = {astro-ph.GA},
       adsurl = {https://ui.adsabs.harvard.edu/abs/2019ApJ...887..235L}
}

@MISC{liu2019a32,
       author = {{Liu}, Daizhong and {A3COSMOS Team}},
        title = "{a3cosmos-gas-evolution: Galaxy cold molecular gas evolution functions}",
 howpublished = {Astrophysics Source Code Library, record ascl:1910.003},
         year = 2019,
        month = oct,
          eid = {ascl:1910.003},
        pages = {ascl:1910.003},
archivePrefix = {ascl},
       eprint = {1910.003},
       adsurl = {https://ui.adsabs.harvard.edu/abs/2019ascl.soft10003L}
}

@ARTICLE{gruppioni2013lf,
       author = {{Gruppioni}, C. and {Pozzi}, F. and {Rodighiero}, G. and {Delvecchio}, I. and {Berta}, S. and {Pozzetti}, L. and {Zamorani}, G. and {Andreani}, P. and {Cimatti}, A. and {Ilbert}, O. and {Le Floc'h}, E. and {Lutz}, D. and {Magnelli}, B. and {Marchetti}, L. and {Monaco}, P. and {Nordon}, R. and {Oliver}, S. and {Popesso}, P. and {Riguccini}, L. and {Roseboom}, I. and {Rosario}, D.~J. and {Sargent}, M. and {Vaccari}, M. and {Altieri}, B. and {Aussel}, H. and {Bongiovanni}, A. and {Cepa}, J. and {Daddi}, E. and {Dom{\'\i}nguez-S{\'a}nchez}, H. and {Elbaz}, D. and {F{\"o}rster Schreiber}, N. and {Genzel}, R. and {Iribarrem}, A. and {Magliocchetti}, M. and {Maiolino}, R. and {Poglitsch}, A. and {P{\'e}rez Garc{\'\i}a}, A. and {Sanchez-Portal}, M. and {Sturm}, E. and {Tacconi}, L. and {Valtchanov}, I. and {Amblard}, A. and {Arumugam}, V. and {Bethermin}, M. and {Bock}, J. and {Boselli}, A. and {Buat}, V. and {Burgarella}, D. and {Castro-Rodr{\'\i}guez}, N. and {Cava}, A. and {Chanial}, P. and {Clements}, D.~L. and {Conley}, A. and {Cooray}, A. and {Dowell}, C.~D. and {Dwek}, E. and {Eales}, S. and {Franceschini}, A. and {Glenn}, J. and {Griffin}, M. and {Hatziminaoglou}, E. and {Ibar}, E. and {Isaak}, K. and {Ivison}, R.~J. and {Lagache}, G. and {Levenson}, L. and {Lu}, N. and {Madden}, S. and {Maffei}, B. and {Mainetti}, G. and {Nguyen}, H.~T. and {O'Halloran}, B. and {Page}, M.~J. and {Panuzzo}, P. and {Papageorgiou}, A. and {Pearson}, C.~P. and {P{\'e}rez-Fournon}, I. and {Pohlen}, M. and {Rigopoulou}, D. and {Rowan-Robinson}, M. and {Schulz}, B. and {Scott}, D. and {Seymour}, N. and {Shupe}, D.~L. and {Smith}, A.~J. and {Stevens}, J.~A. and {Symeonidis}, M. and {Trichas}, M. and {Tugwell}, K.~E. and {Vigroux}, L. and {Wang}, L. and {Wright}, G. and {Xu}, C.~K. and {Zemcov}, M. and {Bardelli}, S. and {Carollo}, M. and {Contini}, T. and {Le F{\'e}vre}, O. and {Lilly}, S. and {Mainieri}, V. and {Renzini}, A. and {Scodeggio}, M. and {Zucca}, E.},
        title = "{The Herschel PEP/HerMES luminosity function - I. Probing the evolution of PACS selected Galaxies to z ≃ 4}",
      journal = {\mnras},
         year = 2013,
        month = jun,
       volume = {432},
       number = {1},
        pages = {23-52},
          doi = {10.1093/mnras/stt308},
archivePrefix = {arXiv},
       eprint = {1302.5209},
 primaryClass = {astro-ph.CO},
       adsurl = {https://ui.adsabs.harvard.edu/abs/2013MNRAS.432...23G}
}

@ARTICLE{scoville2007cosmos,
       author = {{Scoville}, N. and {Aussel}, H. and {Brusa}, M. and {Capak}, P. and {Carollo}, C.~M. and {Elvis}, M. and {Giavalisco}, M. and {Guzzo}, L. and {Hasinger}, G. and {Impey}, C. and {Kneib}, J. -P. and {LeFevre}, O. and {Lilly}, S.~J. and {Mobasher}, B. and {Renzini}, A. and {Rich}, R.~M. and {Sanders}, D.~B. and {Schinnerer}, E. and {Schminovich}, D. and {Shopbell}, P. and {Taniguchi}, Y. and {Tyson}, N.~D.},
        title = "{The Cosmic Evolution Survey (COSMOS): Overview}",
      journal = {\apjs},
         year = 2007,
        month = sep,
       volume = {172},
       number = {1},
        pages = {1-8},
          doi = {10.1086/516585},
archivePrefix = {arXiv},
       eprint = {astro-ph/0612305},
 primaryClass = {astro-ph},
       adsurl = {https://ui.adsabs.harvard.edu/abs/2007ApJS..172....1S}
}

@ARTICLE{chabrier2003imf,
       author = {{Chabrier}, Gilles},
        title = "{Galactic Stellar and Substellar Initial Mass Function}",
      journal = {\pasp},
         year = 2003,
        month = jul,
       volume = {115},
       number = {809},
        pages = {763-795},
          doi = {10.1086/376392},
archivePrefix = {arXiv},
       eprint = {astro-ph/0304382},
 primaryClass = {astro-ph},
       adsurl = {https://ui.adsabs.harvard.edu/abs/2003PASP..115..763C}
}

@ARTICLE{ilbert2013cosmos,
       author = {{Ilbert}, O. and {McCracken}, H.~J. and {Le F{\`e}vre}, O. and {Capak}, P. and {Dunlop}, J. and {Karim}, A. and {Renzini}, M.~A. and {Caputi}, K. and {Boissier}, S. and {Arnouts}, S. and {Aussel}, H. and {Comparat}, J. and {Guo}, Q. and {Hudelot}, P. and {Kartaltepe}, J. and {Kneib}, J.~P. and {Krogager}, J.~K. and {Le Floc'h}, E. and {Lilly}, S. and {Mellier}, Y. and {Milvang-Jensen}, B. and {Moutard}, T. and {Onodera}, M. and {Richard}, J. and {Salvato}, M. and {Sanders}, D.~B. and {Scoville}, N. and {Silverman}, J.~D. and {Taniguchi}, Y. and {Tasca}, L. and {Thomas}, R. and {Toft}, S. and {Tresse}, L. and {Vergani}, D. and {Wolk}, M. and {Zirm}, A.},
        title = "{Mass assembly in quiescent and star-forming galaxies since z ≃ 4 from UltraVISTA}",
      journal = {\aap},
         year = 2013,
        month = aug,
       volume = {556},
          eid = {A55},
        pages = {A55},
          doi = {10.1051/0004-6361/201321100},
archivePrefix = {arXiv},
       eprint = {1301.3157},
 primaryClass = {astro-ph.CO},
       adsurl = {https://ui.adsabs.harvard.edu/abs/2013A&A...556A..55I}
}

@ARTICLE{kennicutt1998sfr,
       author = {{Kennicutt}, Robert C., Jr.},
        title = "{Star Formation in Galaxies Along the Hubble Sequence}",
      journal = {\araa},
         year = 1998,
        month = jan,
       volume = {36},
        pages = {189-232},
          doi = {10.1146/annurev.astro.36.1.189},
archivePrefix = {arXiv},
       eprint = {astro-ph/9807187},
 primaryClass = {astro-ph},
       adsurl = {https://ui.adsabs.harvard.edu/abs/1998ARA&A..36..189K}
}

@ARTICLE{saunders1990modified,
       author = {{Saunders}, W. and {Rowan-Robinson}, M. and {Lawrence}, A. and {Efstathiou}, G. and {Kaiser}, N. and {Ellis}, R.~S. and {Frenk}, C.~S.},
        title = "{The 60-mu.m and far-infrared luminosity functions of IRAS galaxies.}",
      journal = {\mnras},
         year = 1990,
        month = jan,
       volume = {242},
        pages = {318-337},
          doi = {10.1093/mnras/242.3.318},
       adsurl = {https://ui.adsabs.harvard.edu/abs/1990MNRAS.242..318S}
}

@ARTICLE{speagle2014MS,
       author = {{Speagle}, J.~S. and {Steinhardt}, C.~L. and {Capak}, P.~L. and {Silverman}, J.~D.},
        title = "{A Highly Consistent Framework for the Evolution of the Star-Forming ``Main Sequence'' from z \raisebox{-0.5ex}\textasciitilde 0-6}",
      journal = {\apjs},
         year = 2014,
        month = oct,
       volume = {214},
       number = {2},
          eid = {15},
        pages = {15},
          doi = {10.1088/0067-0049/214/2/15},
archivePrefix = {arXiv},
       eprint = {1405.2041},
 primaryClass = {astro-ph.GA},
       adsurl = {https://ui.adsabs.harvard.edu/abs/2014ApJS..214...15S}
}

@ARTICLE{smit2012sfruv,
       author = {{Smit}, Renske and {Bouwens}, Rychard J. and {Franx}, Marijn and {Illingworth}, Garth D. and {Labb{\'e}}, Ivo and {Oesch}, Pascal A. and {van Dokkum}, Pieter G.},
        title = "{The Star Formation Rate Function for Redshift z \raisebox{-0.5ex}\textasciitilde 4-7 Galaxies: Evidence for a Uniform Buildup of Star-forming Galaxies during the First 3 Gyr of Cosmic Time}",
      journal = {\apj},
         year = 2012,
        month = sep,
       volume = {756},
       number = {1},
          eid = {14},
        pages = {14},
          doi = {10.1088/0004-637X/756/1/14},
archivePrefix = {arXiv},
       eprint = {1204.3626},
 primaryClass = {astro-ph.CO},
       adsurl = {https://ui.adsabs.harvard.edu/abs/2012ApJ...756...14S}
}

@ARTICLE{kennicutt1998sfruv,
       author = {{Kennicutt}, Robert C., Jr.},
        title = "{The Global Schmidt Law in Star-forming Galaxies}",
      journal = {\apj},
         year = 1998,
        month = may,
       volume = {498},
       number = {2},
        pages = {541-552},
          doi = {10.1086/305588},
archivePrefix = {arXiv},
       eprint = {astro-ph/9712213},
 primaryClass = {astro-ph},
       adsurl = {https://ui.adsabs.harvard.edu/abs/1998ApJ...498..541K}
}

@ARTICLE{white1991sam,
       author = {{White}, Simon D.~M. and {Frenk}, Carlos S.},
        title = "{Galaxy Formation through Hierarchical Clustering}",
      journal = {\apj},
         year = 1991,
        month = sep,
       volume = {379},
        pages = {52},
          doi = {10.1086/170483},
       adsurl = {https://ui.adsabs.harvard.edu/abs/1991ApJ...379...52W}
}

@ARTICLE{Kauffmann1993sam,
       author = {{Kauffmann}, G. and {White}, S.~D.~M. and {Guiderdoni}, B.},
        title = "{The formation and evolution of galaxies within merging dark matter haloes.}",
      journal = {\mnras},
         year = 1993,
        month = sep,
       volume = {264},
        pages = {201-218},
          doi = {10.1093/mnras/264.1.201},
       adsurl = {https://ui.adsabs.harvard.edu/abs/1993MNRAS.264..201K}
}

@ARTICLE{springel2001sam,
       author = {{Springel}, Volker and {White}, Simon D.~M. and {Tormen}, Giuseppe and {Kauffmann}, Guinevere},
        title = "{Populating a cluster of galaxies - I. Results at [formmu2]z=0}",
      journal = {\mnras},
         year = 2001,
        month = dec,
       volume = {328},
       number = {3},
        pages = {726-750},
          doi = {10.1046/j.1365-8711.2001.04912.x},
archivePrefix = {arXiv},
       eprint = {astro-ph/0012055},
 primaryClass = {astro-ph},
       adsurl = {https://ui.adsabs.harvard.edu/abs/2001MNRAS.328..726S}
}

@ARTICLE{bower2006sam,
       author = {{Bower}, R.~G. and {Benson}, A.~J. and {Malbon}, R. and {Helly}, J.~C. and {Frenk}, C.~S. and {Baugh}, C.~M. and {Cole}, S. and {Lacey}, C.~G.},
        title = "{Breaking the hierarchy of galaxy formation}",
      journal = {\mnras},
         year = 2006,
        month = aug,
       volume = {370},
       number = {2},
        pages = {645-655},
          doi = {10.1111/j.1365-2966.2006.10519.x},
archivePrefix = {arXiv},
       eprint = {astro-ph/0511338},
 primaryClass = {astro-ph},
       adsurl = {https://ui.adsabs.harvard.edu/abs/2006MNRAS.370..645B}
}

@ARTICLE{croton2006sam,
       author = {{Croton}, Darren J. and {Springel}, Volker and {White}, Simon D.~M. and {De Lucia}, G. and {Frenk}, C.~S. and {Gao}, L. and {Jenkins}, A. and {Kauffmann}, G. and {Navarro}, J.~F. and {Yoshida}, N.},
        title = "{The many lives of active galactic nuclei: cooling flows, black holes and the luminosities and colours of galaxies}",
      journal = {\mnras},
         year = 2006,
        month = jan,
       volume = {365},
       number = {1},
        pages = {11-28},
          doi = {10.1111/j.1365-2966.2005.09675.x},
archivePrefix = {arXiv},
       eprint = {astro-ph/0508046},
 primaryClass = {astro-ph},
       adsurl = {https://ui.adsabs.harvard.edu/abs/2006MNRAS.365...11C}
}

@ARTICLE{monaco2007sam,
       author = {{Monaco}, Pierluigi and {Fontanot}, Fabio and {Taffoni}, Giuliano},
        title = "{The MORGANA model for the rise of galaxies and active nuclei}",
      journal = {\mnras},
         year = 2007,
        month = mar,
       volume = {375},
       number = {4},
        pages = {1189-1219},
          doi = {10.1111/j.1365-2966.2006.11253.x},
archivePrefix = {arXiv},
       eprint = {astro-ph/0610805},
 primaryClass = {astro-ph},
       adsurl = {https://ui.adsabs.harvard.edu/abs/2007MNRAS.375.1189M}
}

@ARTICLE{somerville2008sam,
       author = {{Somerville}, Rachel S. and {Hopkins}, Philip F. and {Cox}, Thomas J. and {Robertson}, Brant E. and {Hernquist}, Lars},
        title = "{A semi-analytic model for the co-evolution of galaxies, black holes and active galactic nuclei}",
      journal = {\mnras},
         year = 2008,
        month = dec,
       volume = {391},
       number = {2},
        pages = {481-506},
          doi = {10.1111/j.1365-2966.2008.13805.x},
archivePrefix = {arXiv},
       eprint = {0808.1227},
 primaryClass = {astro-ph},
       adsurl = {https://ui.adsabs.harvard.edu/abs/2008MNRAS.391..481S}
}

@ARTICLE{fontanot2009sam,
       author = {{Fontanot}, Fabio and {De Lucia}, Gabriella and {Monaco}, Pierluigi and {Somerville}, Rachel S. and {Santini}, Paola},
        title = "{The many manifestations of downsizing: hierarchical galaxy formation models confront observations}",
      journal = {\mnras},
         year = 2009,
        month = aug,
       volume = {397},
       number = {4},
        pages = {1776-1790},
          doi = {10.1111/j.1365-2966.2009.15058.x},
archivePrefix = {arXiv},
       eprint = {0901.1130},
 primaryClass = {astro-ph.CO},
       adsurl = {https://ui.adsabs.harvard.edu/abs/2009MNRAS.397.1776F}
}

@ARTICLE{guo2011sam,
       author = {{Guo}, Qi and {White}, Simon and {Boylan-Kolchin}, Michael and {De Lucia}, Gabriella and {Kauffmann}, Guinevere and {Lemson}, Gerard and {Li}, Cheng and {Springel}, Volker and {Weinmann}, Simone},
        title = "{From dwarf spheroidals to cD galaxies: simulating the galaxy population in a {\ensuremath{\Lambda}}CDM cosmology}",
      journal = {\mnras},
         year = 2011,
        month = may,
       volume = {413},
       number = {1},
        pages = {101-131},
          doi = {10.1111/j.1365-2966.2010.18114.x},
archivePrefix = {arXiv},
       eprint = {1006.0106},
 primaryClass = {astro-ph.CO},
       adsurl = {https://ui.adsabs.harvard.edu/abs/2011MNRAS.413..101G}
}

@ARTICLE{benson2012sam,
       author = {{Benson}, Andrew J.},
        title = "{G ALACTICUS: A semi-analytic model of galaxy formation}",
      journal = {\na},
         year = 2012,
        month = feb,
       volume = {17},
       number = {2},
        pages = {175-197},
          doi = {10.1016/j.newast.2011.07.004},
archivePrefix = {arXiv},
       eprint = {1008.1786},
 primaryClass = {astro-ph.CO},
       adsurl = {https://ui.adsabs.harvard.edu/abs/2012NewA...17..175B}
}

@ARTICLE{menci2012sam,
       author = {{Menci}, N. and {Fiore}, F. and {Lamastra}, A.},
        title = "{Galaxy formation in warm dark matter cosmology}",
      journal = {\mnras},
         year = 2012,
        month = apr,
       volume = {421},
       number = {3},
        pages = {2384-2394},
          doi = {10.1111/j.1365-2966.2012.20470.x},
archivePrefix = {arXiv},
       eprint = {1201.1617},
 primaryClass = {astro-ph.CO},
       adsurl = {https://ui.adsabs.harvard.edu/abs/2012MNRAS.421.2384M}
}

@ARTICLE{somerville2012sam,
       author = {{Somerville}, Rachel S. and {Gilmore}, Rudy C. and {Primack}, Joel R. and {Dom{\'\i}nguez}, Alberto},
        title = "{Galaxy properties from the ultraviolet to the far-infrared: {\ensuremath{\Lambda}} cold dark matter models confront observations}",
      journal = {\mnras},
         year = 2012,
        month = jul,
       volume = {423},
       number = {3},
        pages = {1992-2015},
          doi = {10.1111/j.1365-2966.2012.20490.x},
archivePrefix = {arXiv},
       eprint = {1104.0669},
 primaryClass = {astro-ph.CO},
       adsurl = {https://ui.adsabs.harvard.edu/abs/2012MNRAS.423.1992S}
}

@ARTICLE{henriques2013sam,
       author = {{Henriques}, Bruno M.~B. and {White}, Simon D.~M. and {Thomas}, Peter A. and {Angulo}, Raul E. and {Guo}, Qi and {Lemson}, Gerard and {Springel}, Volker},
        title = "{Simulations of the galaxy population constrained by observations from z = 3 to the present day: implications for galactic winds and the fate of their ejecta}",
      journal = {\mnras},
         year = 2013,
        month = jun,
       volume = {431},
       number = {4},
        pages = {3373-3395},
          doi = {10.1093/mnras/stt415},
archivePrefix = {arXiv},
       eprint = {1212.1717},
 primaryClass = {astro-ph.CO},
       adsurl = {https://ui.adsabs.harvard.edu/abs/2013MNRAS.431.3373H}
}

@ARTICLE{gruppioni2015sam,
       author = {{Gruppioni}, C. and {Calura}, F. and {Pozzi}, F. and {Delvecchio}, I. and {Berta}, S. and {De Lucia}, G. and {Fontanot}, F. and {Franceschini}, A. and {Marchetti}, L. and {Menci}, N. and {Monaco}, P. and {Vaccari}, M.},
        title = "{Star formation in Herschel's Monsters versus semi-analytic models}",
      journal = {\mnras},
         year = 2015,
        month = aug,
       volume = {451},
       number = {4},
        pages = {3419-3426},
          doi = {10.1093/mnras/stv1204},
archivePrefix = {arXiv},
       eprint = {1506.01518},
 primaryClass = {astro-ph.GA},
       adsurl = {https://ui.adsabs.harvard.edu/abs/2015MNRAS.451.3419G}
}

@ARTICLE{henriques2015sam,
       author = {{Henriques}, Bruno M.~B. and {White}, Simon D.~M. and {Thomas}, Peter A. and {Angulo}, Raul and {Guo}, Qi and {Lemson}, Gerard and {Springel}, Volker and {Overzier}, Roderik},
        title = "{Galaxy formation in the Planck cosmology - I. Matching the observed evolution of star formation rates, colours and stellar masses}",
      journal = {\mnras},
         year = 2015,
        month = aug,
       volume = {451},
       number = {3},
        pages = {2663-2680},
          doi = {10.1093/mnras/stv705},
archivePrefix = {arXiv},
       eprint = {1410.0365},
 primaryClass = {astro-ph.GA},
       adsurl = {https://ui.adsabs.harvard.edu/abs/2015MNRAS.451.2663H}
}

@ARTICLE{reddy2008sfrf,
       author = {{Reddy}, Naveen A. and {Steidel}, Charles C. and {Pettini}, Max and {Adelberger}, Kurt L. and {Shapley}, Alice E. and {Erb}, Dawn K. and {Dickinson}, Mark},
        title = "{Multiwavelength Constraints on the Cosmic Star Formation History from Spectroscopy: The Rest-Frame Ultraviolet, H{\ensuremath{\alpha}}, and Infrared Luminosity Functions at Redshifts 1.9 lesssim z lesssim 3.4}",
      journal = {\apjs},
         year = 2008,
        month = mar,
       volume = {175},
       number = {1},
        pages = {48-85},
          doi = {10.1086/521105},
archivePrefix = {arXiv},
       eprint = {0706.4091},
 primaryClass = {astro-ph},
       adsurl = {https://ui.adsabs.harvard.edu/abs/2008ApJS..175...48R}
}

@ARTICLE{ly2011sfrf,
       author = {{Ly}, Chun and {Lee}, Janice C. and {Dale}, Daniel A. and {Momcheva}, Ivelina and {Salim}, Samir and {Staudaher}, Shawn and {Moore}, Carolynn A. and {Finn}, Rose},
        title = "{The H{\ensuremath{\alpha}} Luminosity Function and Star Formation Rate Volume Density at z = 0.8 from the NEWFIRM H{\ensuremath{\alpha}} Survey}",
      journal = {\apj},
         year = 2011,
        month = jan,
       volume = {726},
       number = {2},
          eid = {109},
        pages = {109},
          doi = {10.1088/0004-637X/726/2/109},
archivePrefix = {arXiv},
       eprint = {1011.2759},
 primaryClass = {astro-ph.CO},
       adsurl = {https://ui.adsabs.harvard.edu/abs/2011ApJ...726..109L}
}

@ARTICLE{magnelli2011sfrf,
       author = {{Magnelli}, B. and {Elbaz}, D. and {Chary}, R.~R. and {Dickinson}, M. and {Le Borgne}, D. and {Frayer}, D.~T. and {Willmer}, C.~N.~A.},
        title = "{Evolution of the dusty infrared luminosity function from z = 0 to z = 2.3 using observations from Spitzer}",
      journal = {\aap},
         year = 2011,
        month = apr,
       volume = {528},
          eid = {A35},
        pages = {A35},
          doi = {10.1051/0004-6361/200913941},
archivePrefix = {arXiv},
       eprint = {1101.2467},
 primaryClass = {astro-ph.CO},
       adsurl = {https://ui.adsabs.harvard.edu/abs/2011A&A...528A..35M}
}

@ARTICLE{patel2013sfrf,
       author = {{Patel}, Shannon G. and {van Dokkum}, Pieter G. and {Franx}, Marijn and {Quadri}, Ryan F. and {Muzzin}, Adam and {Marchesini}, Danilo and {Williams}, Rik J. and {Holden}, Bradford P. and {Stefanon}, Mauro},
        title = "{HST/WFC3 Confirmation of the Inside-out Growth of Massive Galaxies at 0 < z < 2 and Identification of Their Star-forming Progenitors at z \raisebox{-0.5ex}\textasciitilde 3}",
      journal = {\apj},
         year = 2013,
        month = mar,
       volume = {766},
       number = {1},
          eid = {15},
        pages = {15},
          doi = {10.1088/0004-637X/766/1/15},
archivePrefix = {arXiv},
       eprint = {1208.0341},
 primaryClass = {astro-ph.CO},
       adsurl = {https://ui.adsabs.harvard.edu/abs/2013ApJ...766...15P}
}

@ARTICLE{sobral2013sfrf,
       author = {{Sobral}, David and {Smail}, Ian and {Best}, Philip N. and {Geach}, James E. and {Matsuda}, Yuichi and {Stott}, John P. and {Cirasuolo}, Michele and {Kurk}, Jaron},
        title = "{A large H{\ensuremath{\alpha}} survey at z = 2.23, 1.47, 0.84 and 0.40: the 11 Gyr evolution of star-forming galaxies from HiZELS★}",
      journal = {\mnras},
         year = 2013,
        month = jan,
       volume = {428},
       number = {2},
        pages = {1128-1146},
          doi = {10.1093/mnras/sts096},
archivePrefix = {arXiv},
       eprint = {1202.3436},
 primaryClass = {astro-ph.CO},
       adsurl = {https://ui.adsabs.harvard.edu/abs/2013MNRAS.428.1128S}
}

@ARTICLE{duncan2014sfrf,
       author = {{Duncan}, K. and {Conselice}, C.~J. and {Mortlock}, A. and {Hartley}, W.~G. and {Guo}, Y. and {Ferguson}, H.~C. and {Dav{\'e}}, R. and {Lu}, Y. and {Ownsworth}, J. and {Ashby}, M.~L.~N. and {Dekel}, A. and {Dickinson}, M. and {Faber}, S. and {Giavalisco}, M. and {Grogin}, N. and {Kocevski}, D. and {Koekemoer}, A. and {Somerville}, R.~S. and {White}, C.~E.},
        title = "{The mass evolution of the first galaxies: stellar mass functions and star formation rates at 4 < z < 7 in the CANDELS GOODS-South field}",
      journal = {\mnras},
         year = 2014,
        month = nov,
       volume = {444},
       number = {3},
        pages = {2960-2984},
          doi = {10.1093/mnras/stu1622},
archivePrefix = {arXiv},
       eprint = {1408.2527},
 primaryClass = {astro-ph.GA},
       adsurl = {https://ui.adsabs.harvard.edu/abs/2014MNRAS.444.2960D}
}

@ARTICLE{bouwens2015sfrf,
       author = {{Bouwens}, R.~J. and {Illingworth}, G.~D. and {Oesch}, P.~A. and {Trenti}, M. and {Labb{\'e}}, I. and {Bradley}, L. and {Carollo}, M. and {van Dokkum}, P.~G. and {Gonzalez}, V. and {Holwerda}, B. and {Franx}, M. and {Spitler}, L. and {Smit}, R. and {Magee}, D.},
        title = "{UV Luminosity Functions at Redshifts z {\ensuremath{\sim}} 4 to z {\ensuremath{\sim}} 10: 10,000 Galaxies from HST Legacy Fields}",
      journal = {\apj},
         year = 2015,
        month = apr,
       volume = {803},
       number = {1},
          eid = {34},
        pages = {34},
          doi = {10.1088/0004-637X/803/1/34},
archivePrefix = {arXiv},
       eprint = {1403.4295},
 primaryClass = {astro-ph.CO},
       adsurl = {https://ui.adsabs.harvard.edu/abs/2015ApJ...803...34B}
}

@ARTICLE{alavi2016sfrf,
       author = {{Alavi}, Anahita and {Siana}, Brian and {Richard}, Johan and {Rafelski}, Marc and {Jauzac}, Mathilde and {Limousin}, Marceau and {Freeman}, William R. and {Scarlata}, Claudia and {Robertson}, Brant and {Stark}, Daniel P. and {Teplitz}, Harry I. and {Desai}, Vandana},
        title = "{The Evolution of the Faint End of the UV Luminosity Function during the Peak Epoch of Star Formation (1 < z < 3)}",
      journal = {\apj},
         year = 2016,
        month = nov,
       volume = {832},
       number = {1},
          eid = {56},
        pages = {56},
          doi = {10.3847/0004-637X/832/1/56},
archivePrefix = {arXiv},
       eprint = {1606.00469},
 primaryClass = {astro-ph.GA},
       adsurl = {https://ui.adsabs.harvard.edu/abs/2016ApJ...832...56A}
}

@ARTICLE{parsa2016sfrf,
       author = {{Parsa}, Shaghayegh and {Dunlop}, James S. and {McLure}, Ross J. and {Mortlock}, Alice},
        title = "{The galaxy UV luminosity function at z ≃ 2-4; new results on faint-end slope and the evolution of luminosity density}",
      journal = {\mnras},
         year = 2016,
        month = mar,
       volume = {456},
       number = {3},
        pages = {3194-3211},
          doi = {10.1093/mnras/stv2857},
archivePrefix = {arXiv},
       eprint = {1507.05629},
 primaryClass = {astro-ph.GA},
       adsurl = {https://ui.adsabs.harvard.edu/abs/2016MNRAS.456.3194P}
}

@ARTICLE{cucciati2012sfrd,
       author = {{Cucciati}, O. and {Tresse}, L. and {Ilbert}, O. and {Le F{\`e}vre}, O. and {Garilli}, B. and {Le Brun}, V. and {Cassata}, P. and {Franzetti}, P. and {Maccagni}, D. and {Scodeggio}, M. and {Zucca}, E. and {Zamorani}, G. and {Bardelli}, S. and {Bolzonella}, M. and {Bielby}, R.~M. and {McCracken}, H.~J. and {Zanichelli}, A. and {Vergani}, D.},
        title = "{The star formation rate density and dust attenuation evolution over 12 Gyr with the VVDS surveys}",
      journal = {\aap},
         year = 2012,
        month = mar,
       volume = {539},
          eid = {A31},
        pages = {A31},
          doi = {10.1051/0004-6361/201118010},
archivePrefix = {arXiv},
       eprint = {1109.1005},
 primaryClass = {astro-ph.CO},
       adsurl = {https://ui.adsabs.harvard.edu/abs/2012A&A...539A..31C}
}

@ARTICLE{oesch2010sfrf,
       author = {{Oesch}, P.~A. and {Bouwens}, R.~J. and {Carollo}, C.~M. and {Illingworth}, G.~D. and {Magee}, D. and {Trenti}, M. and {Stiavelli}, M. and {Franx}, M. and {Labb{\'e}}, I. and {van Dokkum}, P.~G.},
        title = "{The Evolution of the Ultraviolet Luminosity Function from z \raisebox{-0.5ex}\textasciitilde 0.75 to z \raisebox{-0.5ex}\textasciitilde 2.5 Using HST ERS WFC3/UVIS Observations}",
      journal = {\apjl},
         year = 2010,
        month = dec,
       volume = {725},
       number = {2},
        pages = {L150-L155},
          doi = {10.1088/2041-8205/725/2/L150},
archivePrefix = {arXiv},
       eprint = {1005.1661},
 primaryClass = {astro-ph.CO},
       adsurl = {https://ui.adsabs.harvard.edu/abs/2010ApJ...725L.150O}
}

@ARTICLE{vander2010sfrf,
       author = {{van der Burg}, R.~F.~J. and {Hildebrandt}, H. and {Erben}, T.},
        title = "{The UV galaxy luminosity function at z = 3-5 from the CFHT Legacy Survey Deep fields}",
      journal = {\aap},
         year = 2010,
        month = nov,
       volume = {523},
          eid = {A74},
        pages = {A74},
          doi = {10.1051/0004-6361/200913812},
archivePrefix = {arXiv},
       eprint = {1009.0758},
 primaryClass = {astro-ph.CO},
       adsurl = {https://ui.adsabs.harvard.edu/abs/2010A&A...523A..74V}
}

@ARTICLE{dave2011simul,
       author = {{Dav{\'e}}, Romeel and {Oppenheimer}, Benjamin D. and {Finlator}, Kristian},
        title = "{Galaxy evolution in cosmological simulations with outflows - I. Stellar masses and star formation rates}",
      journal = {\mnras},
         year = 2011,
        month = jul,
       volume = {415},
       number = {1},
        pages = {11-31},
          doi = {10.1111/j.1365-2966.2011.18680.x},
archivePrefix = {arXiv},
       eprint = {1103.3528},
 primaryClass = {astro-ph.CO},
       adsurl = {https://ui.adsabs.harvard.edu/abs/2011MNRAS.415...11D}
}

@ARTICLE{katsianis2021sfrd,
       author = {{Katsianis}, Antonios and {Yang}, Xiaohu and {Zheng}, Xianzhong},
        title = "{The Observed Cosmic Star Formation Rate Density Has an Evolution that Resembles a {\ensuremath{\Gamma}}(a, bt) Distribution and Can Be Described Successfully by Only Two Parameters}",
      journal = {\apj},
         year = 2021,
        month = oct,
       volume = {919},
       number = {2},
          eid = {88},
        pages = {88},
          doi = {10.3847/1538-4357/ac11f2},
archivePrefix = {arXiv},
       eprint = {2107.02733},
 primaryClass = {astro-ph.GA},
       adsurl = {https://ui.adsabs.harvard.edu/abs/2021ApJ...919...88K}
}

@ARTICLE{katsianis2017eagle,
       author = {{Katsianis}, A. and {Blanc}, G. and {Lagos}, C.~P. and {Tejos}, N. and {Bower}, R.~G. and {Alavi}, A. and {Gonzalez}, V. and {Theuns}, T. and {Schaller}, M. and {Lopez}, S.},
        title = "{The evolution of the star formation rate function in the EAGLE simulations: a comparison with UV, IR and H{\ensuremath{\alpha}} observations from z {\ensuremath{\sim}} 8 to z {\ensuremath{\sim}} 0}",
      journal = {\mnras},
         year = 2017,
        month = nov,
       volume = {472},
       number = {1},
        pages = {919-939},
          doi = {10.1093/mnras/stx2020},
archivePrefix = {arXiv},
       eprint = {1708.01913},
 primaryClass = {astro-ph.GA},
       adsurl = {https://ui.adsabs.harvard.edu/abs/2017MNRAS.472..919K}
}

@ARTICLE{tescari2014simul,
       author = {{Tescari}, E. and {Katsianis}, A. and {Wyithe}, J.~S.~B. and {Dolag}, K. and {Tornatore}, L. and {Barai}, P. and {Viel}, M. and {Borgani}, S.},
        title = "{Simulated star formation rate functions at z {\ensuremath{\sim}} 4-7, and the role of feedback in high-z galaxies}",
      journal = {\mnras},
         year = 2014,
        month = mar,
       volume = {438},
       number = {4},
        pages = {3490-3506},
          doi = {10.1093/mnras/stt2461},
archivePrefix = {arXiv},
       eprint = {1312.5310},
 primaryClass = {astro-ph.CO},
       adsurl = {https://ui.adsabs.harvard.edu/abs/2014MNRAS.438.3490T}
}

@ARTICLE{fontanot2012sam,
       author = {{Fontanot}, Fabio and {Springel}, Volker and {Angulo}, Raul E. and {Henriques}, Bruno},
        title = "{Semi-analytic galaxy formation in early dark energy cosmologies}",
      journal = {\mnras},
         year = 2012,
        month = nov,
       volume = {426},
       number = {3},
        pages = {2335-2341},
          doi = {10.1111/j.1365-2966.2012.21880.x},
archivePrefix = {arXiv},
       eprint = {1207.1723},
 primaryClass = {astro-ph.CO},
       adsurl = {https://ui.adsabs.harvard.edu/abs/2012MNRAS.426.2335F}
}

@ARTICLE{ono2018uvLF,
       author = {{Ono}, Yoshiaki and {Ouchi}, Masami and {Harikane}, Yuichi and {Toshikawa}, Jun and {Rauch}, Michael and {Yuma}, Suraphong and {Sawicki}, Marcin and {Shibuya}, Takatoshi and {Shimasaku}, Kazuhiro and {Oguri}, Masamune and {Willott}, Chris and {Akhlaghi}, Mohammad and {Akiyama}, Masayuki and {Coupon}, Jean and {Kashikawa}, Nobunari and {Komiyama}, Yutaka and {Konno}, Akira and {Lin}, Lihwai and {Matsuoka}, Yoshiki and {Miyazaki}, Satoshi and {Nagao}, Tohru and {Nakajima}, Kimihiko and {Silverman}, John and {Tanaka}, Masayuki and {Taniguchi}, Yoshiaki and {Wang}, Shiang-Yu},
        title = "{Great Optically Luminous Dropout Research Using Subaru HSC (GOLDRUSH). I. UV luminosity functions at z {\ensuremath{\sim}} 4-7 derived with the half-million dropouts on the 100 deg$^{2}$ sky}",
      journal = {\pasj},
         year = 2018,
        month = jan,
       volume = {70},
          eid = {S10},
        pages = {S10},
          doi = {10.1093/pasj/psx103},
archivePrefix = {arXiv},
       eprint = {1704.06004},
 primaryClass = {astro-ph.GA},
       adsurl = {https://ui.adsabs.harvard.edu/abs/2018PASJ...70S..10O}
}

@ARTICLE{mehta2017UV_LF,
       author = {{Mehta}, Vihang and {Scarlata}, Claudia and {Rafelski}, Marc and {Gburek}, Timothy and {Teplitz}, Harry I. and {Alavi}, Anahita and {Boylan-Kolchin}, Michael and {Finkelstein}, Steven and {Gardner}, Jonathan P. and {Grogin}, Norman and {Koekemoer}, Anton and {Kurczynski}, Peter and {Siana}, Brian and {Codoreanu}, Alex and {de Mello}, Duilia F. and {Lee}, Kyoung-Soo and {Soto}, Emmaris},
        title = "{UVUDF: UV Luminosity Functions at the Cosmic High Noon}",
      journal = {\apj},
         year = 2017,
        month = mar,
       volume = {838},
       number = {1},
          eid = {29},
        pages = {29},
          doi = {10.3847/1538-4357/aa6259},
archivePrefix = {arXiv},
       eprint = {1702.06953},
 primaryClass = {astro-ph.GA},
       adsurl = {https://ui.adsabs.harvard.edu/abs/2017ApJ...838...29M}
}

@ARTICLE{adams2020uvlf,
       author = {{Adams}, N.~J. and {Bowler}, R.~A.~A. and {Jarvis}, M.~J. and {H{\"a}u{\ss}ler}, B. and {McLure}, R.~J. and {Bunker}, A. and {Dunlop}, J.~S. and {Verma}, A.},
        title = "{The rest-frame UV luminosity function at z ≃ 4: a significant contribution of AGNs to the bright end of the galaxy population}",
      journal = {\mnras},
         year = 2020,
        month = may,
       volume = {494},
       number = {2},
        pages = {1771-1783},
          doi = {10.1093/mnras/staa687},
archivePrefix = {arXiv},
       eprint = {1912.01626},
 primaryClass = {astro-ph.GA},
       adsurl = {https://ui.adsabs.harvard.edu/abs/2020MNRAS.494.1771A}
}

@ARTICLE{bouwens2021UV_LF,
       author = {{Bouwens}, R.~J. and {Oesch}, P.~A. and {Stefanon}, M. and {Illingworth}, G. and {Labb{\'e}}, I. and {Reddy}, N. and {Atek}, H. and {Montes}, M. and {Naidu}, R. and {Nanayakkara}, T. and {Nelson}, E. and {Wilkins}, S.},
        title = "{New Determinations of the UV Luminosity Functions from z   9 to 2 Show a Remarkable Consistency with Halo Growth and a Constant Star Formation Efficiency}",
      journal = {\aj},
         year = 2021,
        month = aug,
       volume = {162},
       number = {2},
          eid = {47},
        pages = {47},
          doi = {10.3847/1538-3881/abf83e},
archivePrefix = {arXiv},
       eprint = {2102.07775},
 primaryClass = {astro-ph.GA},
       adsurl = {https://ui.adsabs.harvard.edu/abs/2021AJ....162...47B}
}

@ARTICLE{bouwens2022uvlf,
       author = {{Bouwens}, R.~J. and {Illingworth}, G. and {Ellis}, R.~S. and {Oesch}, P. and {Stefanon}, M.},
        title = "{z   2-9 Galaxies Magnified by the Hubble Frontier Field Clusters. II. Luminosity Functions and Constraints on a Faint-end Turnover}",
      journal = {\apj},
         year = 2022,
        month = nov,
       volume = {940},
       number = {1},
          eid = {55},
        pages = {55},
          doi = {10.3847/1538-4357/ac86d1},
archivePrefix = {arXiv},
       eprint = {2205.11526},
 primaryClass = {astro-ph.GA},
       adsurl = {https://ui.adsabs.harvard.edu/abs/2022ApJ...940...55B}
}

@ARTICLE{hao2011sfr,
       author = {{Hao}, Cai-Na and {Kennicutt}, Robert C. and {Johnson}, Benjamin D. and {Calzetti}, Daniela and {Dale}, Daniel A. and {Moustakas}, John},
        title = "{Dust-corrected Star Formation Rates of Galaxies. II. Combinations of Ultraviolet and Infrared Tracers}",
      journal = {\apj},
         year = 2011,
        month = nov,
       volume = {741},
       number = {2},
          eid = {124},
        pages = {124},
          doi = {10.1088/0004-637X/741/2/124},
archivePrefix = {arXiv},
       eprint = {1108.2837},
 primaryClass = {astro-ph.CO},
       adsurl = {https://ui.adsabs.harvard.edu/abs/2011ApJ...741..124H}
}

@ARTICLE{bouwens2012sfr,
       author = {{Bouwens}, R.~J. and {Illingworth}, G.~D. and {Oesch}, P.~A. and {Franx}, M. and {Labb{\'e}}, I. and {Trenti}, M. and {van Dokkum}, P. and {Carollo}, C.~M. and {Gonz{\'a}lez}, V. and {Smit}, R. and {Magee}, D.},
        title = "{UV-continuum Slopes at z \raisebox{-0.5ex}\textasciitilde 4-7 from the HUDF09+ERS+CANDELS Observations: Discovery of a Well-defined UV Color-Magnitude Relationship for z >= 4 Star-forming Galaxies}",
      journal = {\apj},
         year = 2012,
        month = aug,
       volume = {754},
       number = {2},
          eid = {83},
        pages = {83},
          doi = {10.1088/0004-637X/754/2/83},
archivePrefix = {arXiv},
       eprint = {1109.0994},
 primaryClass = {astro-ph.CO},
       adsurl = {https://ui.adsabs.harvard.edu/abs/2012ApJ...754...83B}
}

@ARTICLE{tacchella2013sfr,
       author = {{Tacchella}, Sandro and {Trenti}, Michele and {Carollo}, C. Marcella},
        title = "{A Physical Model for the 0 <\raisebox{-0.5ex}\textasciitilde z <\raisebox{-0.5ex}\textasciitilde 8 Redshift Evolution of the Galaxy Ultraviolet Luminosity and Stellar Mass Functions}",
      journal = {\apjl},
         year = 2013,
        month = may,
       volume = {768},
       number = {2},
          eid = {L37},
        pages = {L37},
          doi = {10.1088/2041-8205/768/2/L37},
archivePrefix = {arXiv},
       eprint = {1211.2825},
 primaryClass = {astro-ph.CO},
       adsurl = {https://ui.adsabs.harvard.edu/abs/2013ApJ...768L..37T}
}

@ARTICLE{dallavecchia2008sn,
       author = {{Dalla Vecchia}, Claudio and {Schaye}, Joop},
        title = "{Simulating galactic outflows with kinetic supernova feedback}",
      journal = {\mnras},
         year = 2008,
        month = jul,
       volume = {387},
       number = {4},
        pages = {1431-1444},
          doi = {10.1111/j.1365-2966.2008.13322.x},
archivePrefix = {arXiv},
       eprint = {0801.2770},
 primaryClass = {astro-ph},
       adsurl = {https://ui.adsabs.harvard.edu/abs/2008MNRAS.387.1431D}
}

@ARTICLE{Hirschmann2016sam,
       author = {{Hirschmann}, Michaela and {De Lucia}, Gabriella and {Fontanot}, Fabio},
        title = "{Galaxy assembly, stellar feedback and metal enrichment: the view from the GAEA model}",
      journal = {\mnras},
         year = 2016,
        month = sep,
       volume = {461},
       number = {2},
        pages = {1760-1785},
          doi = {10.1093/mnras/stw1318},
archivePrefix = {arXiv},
       eprint = {1512.04531},
 primaryClass = {astro-ph.GA},
       adsurl = {https://ui.adsabs.harvard.edu/abs/2016MNRAS.461.1760H}
}

@ARTICLE{fontanot2020sam,
       author = {{Fontanot}, Fabio and {De Lucia}, Gabriella and {Hirschmann}, Michaela and {Xie}, Lizhi and {Monaco}, Pierluigi and {Menci}, Nicola and {Fiore}, Fabrizio and {Feruglio}, Chiara and {Cristiani}, Stefano and {Shankar}, Francesco},
        title = "{The rise of active galactic nuclei in the galaxy evolution and assembly semi-analytic model}",
      journal = {\mnras},
         year = 2020,
        month = aug,
       volume = {496},
       number = {3},
        pages = {3943-3960},
          doi = {10.1093/mnras/staa1716},
archivePrefix = {arXiv},
       eprint = {2002.10576},
 primaryClass = {astro-ph.CO},
       adsurl = {https://ui.adsabs.harvard.edu/abs/2020MNRAS.496.3943F}
}

@ARTICLE{schechter1976,
       author = {{Schechter}, P.},
        title = "{An analytic expression for the luminosity function for galaxies.}",
      journal = {\apj},
         year = 1976,
        month = jan,
       volume = {203},
        pages = {297-306},
          doi = {10.1086/154079},
       adsurl = {https://ui.adsabs.harvard.edu/abs/1976ApJ...203..297S}
}

@ARTICLE{caputi2007lfevol,
       author = {{Caputi}, K.~I. and {Lagache}, G. and {Yan}, Lin and {Dole}, H. and {Bavouzet}, N. and {Le Floc'h}, E. and {Choi}, P.~I. and {Helou}, G. and {Reddy}, N.},
        title = "{The Infrared Luminosity Function of Galaxies at Redshifts z = 1 and z \raisebox{-0.5ex}\textasciitilde 2 in the GOODS Fields}",
      journal = {\apj},
         year = 2007,
        month = may,
       volume = {660},
       number = {1},
        pages = {97-116},
          doi = {10.1086/512667},
archivePrefix = {arXiv},
       eprint = {astro-ph/0701283},
 primaryClass = {astro-ph},
       adsurl = {https://ui.adsabs.harvard.edu/abs/2007ApJ...660...97C}
}

@ARTICLE{bethermin2011lfevol,
       author = {{B{\'e}thermin}, M. and {Dole}, H. and {Lagache}, G. and {Le Borgne}, D. and {Penin}, A.},
        title = "{Modeling the evolution of infrared galaxies: a parametric backward evolution model}",
      journal = {\aap},
         year = 2011,
        month = may,
       volume = {529},
          eid = {A4},
        pages = {A4},
          doi = {10.1051/0004-6361/201015841},
archivePrefix = {arXiv},
       eprint = {1010.1150},
 primaryClass = {astro-ph.CO},
       adsurl = {https://ui.adsabs.harvard.edu/abs/2011A&A...529A...4B}
}

@ARTICLE{parente2023dust_sam,
       author = {{Parente}, Massimiliano and {Ragone-Figueroa}, Cinthia and {Granato}, Gian Luigi and {Lapi}, Andrea},
        title = "{The z {\ensuremath{\lesssim}} 1 drop of cosmic dust abundance in a semi-analytic framework}",
      journal = {\mnras},
         year = 2023,
        month = jun,
       volume = {521},
       number = {4},
        pages = {6105-6123},
          doi = {10.1093/mnras/stad907},
archivePrefix = {arXiv},
       eprint = {2302.03058},
 primaryClass = {astro-ph.GA},
       adsurl = {https://ui.adsabs.harvard.edu/abs/2023MNRAS.521.6105P}
}

@ARTICLE{pillepich2018illustrisTNG,
       author = {{Pillepich}, Annalisa and {Springel}, Volker and {Nelson}, Dylan and {Genel}, Shy and {Naiman}, Jill and {Pakmor}, R{\"u}diger and {Hernquist}, Lars and {Torrey}, Paul and {Vogelsberger}, Mark and {Weinberger}, Rainer and {Marinacci}, Federico},
        title = "{Simulating galaxy formation with the IllustrisTNG model}",
      journal = {\mnras},
         year = 2018,
        month = jan,
       volume = {473},
       number = {3},
        pages = {4077-4106},
          doi = {10.1093/mnras/stx2656},
archivePrefix = {arXiv},
       eprint = {1703.02970},
 primaryClass = {astro-ph.GA},
       adsurl = {https://ui.adsabs.harvard.edu/abs/2018MNRAS.473.4077P}
}

@ARTICLE{genel2014illustris,
       author = {{Genel}, Shy and {Vogelsberger}, Mark and {Springel}, Volker and {Sijacki}, Debora and {Nelson}, Dylan and {Snyder}, Greg and {Rodriguez-Gomez}, Vicente and {Torrey}, Paul and {Hernquist}, Lars},
        title = "{Introducing the Illustris project: the evolution of galaxy populations across cosmic time}",
      journal = {\mnras},
         year = 2014,
        month = nov,
       volume = {445},
       number = {1},
        pages = {175-200},
          doi = {10.1093/mnras/stu1654},
archivePrefix = {arXiv},
       eprint = {1405.3749},
 primaryClass = {astro-ph.CO},
       adsurl = {https://ui.adsabs.harvard.edu/abs/2014MNRAS.445..175G}
}

@ARTICLE{vogelsberger2014illustris,
       author = {{Vogelsberger}, Mark and {Genel}, Shy and {Springel}, Volker and {Torrey}, Paul and {Sijacki}, Debora and {Xu}, Dandan and {Snyder}, Greg and {Nelson}, Dylan and {Hernquist}, Lars},
        title = "{Introducing the Illustris Project: simulating the coevolution of dark and visible matter in the Universe}",
      journal = {\mnras},
         year = 2014,
        month = oct,
       volume = {444},
       number = {2},
        pages = {1518-1547},
          doi = {10.1093/mnras/stu1536},
archivePrefix = {arXiv},
       eprint = {1405.2921},
 primaryClass = {astro-ph.CO},
       adsurl = {https://ui.adsabs.harvard.edu/abs/2014MNRAS.444.1518V}
}

@ARTICLE{springel2010arepo,
       author = {{Springel}, Volker},
        title = "{E pur si muove: Galilean-invariant cosmological hydrodynamical simulations on a moving mesh}",
      journal = {\mnras},
         year = 2010,
        month = jan,
       volume = {401},
       number = {2},
        pages = {791-851},
          doi = {10.1111/j.1365-2966.2009.15715.x},
archivePrefix = {arXiv},
       eprint = {0901.4107},
 primaryClass = {astro-ph.CO},
       adsurl = {https://ui.adsabs.harvard.edu/abs/2010MNRAS.401..791S}
}

@ARTICLE{dave2017mufasa,
       author = {{Dav{\'e}}, Romeel and {Rafieferantsoa}, Mika H. and {Thompson}, Robert J. and {Hopkins}, Philip F.},
        title = "{MUFASA: Galaxy star formation, gas, and metal properties across cosmic time}",
      journal = {\mnras},
         year = 2017,
        month = may,
       volume = {467},
       number = {1},
        pages = {115-132},
          doi = {10.1093/mnras/stx108},
archivePrefix = {arXiv},
       eprint = {1610.01626},
 primaryClass = {astro-ph.GA},
       adsurl = {https://ui.adsabs.harvard.edu/abs/2017MNRAS.467..115D}
}

@ARTICLE{hopking2015gizmo,
       author = {{Hopkins}, Philip F.},
        title = "{A new class of accurate, mesh-free hydrodynamic simulation methods}",
      journal = {\mnras},
         year = 2015,
        month = jun,
       volume = {450},
       number = {1},
        pages = {53-110},
          doi = {10.1093/mnras/stv195},
archivePrefix = {arXiv},
       eprint = {1409.7395},
 primaryClass = {astro-ph.CO},
       adsurl = {https://ui.adsabs.harvard.edu/abs/2015MNRAS.450...53H}
}

@ARTICLE{dave2019simba,
       author = {{Dav{\'e}}, Romeel and {Angl{\'e}s-Alc{\'a}zar}, Daniel and {Narayanan}, Desika and {Li}, Qi and {Rafieferantsoa}, Mika H. and {Appleby}, Sarah},
        title = "{SIMBA: Cosmological simulations with black hole growth and feedback}",
      journal = {\mnras},
         year = 2019,
        month = jun,
       volume = {486},
       number = {2},
        pages = {2827-2849},
          doi = {10.1093/mnras/stz937},
archivePrefix = {arXiv},
       eprint = {1901.10203},
 primaryClass = {astro-ph.GA},
       adsurl = {https://ui.adsabs.harvard.edu/abs/2019MNRAS.486.2827D}
}

@ARTICLE{henriques2020lgalaxies,
       author = {{Henriques}, Bruno M.~B. and {Yates}, Robert M. and {Fu}, Jian and {Guo}, Qi and {Kauffmann}, Guinevere and {Srisawat}, Chaichalit and {Thomas}, Peter A. and {White}, Simon D.~M.},
        title = "{L-GALAXIES 2020: Spatially resolved cold gas phases, star formation, and chemical enrichment in galactic discs}",
      journal = {\mnras},
         year = 2020,
        month = feb,
       volume = {491},
       number = {4},
        pages = {5795-5814},
          doi = {10.1093/mnras/stz3233},
archivePrefix = {arXiv},
       eprint = {2003.05944},
 primaryClass = {astro-ph.GA},
       adsurl = {https://ui.adsabs.harvard.edu/abs/2020MNRAS.491.5795H}
}

@ARTICLE{kennicutt2012sfr,
       author = {{Kennicutt}, Robert C. and {Evans}, Neal J.},
        title = "{Star Formation in the Milky Way and Nearby Galaxies}",
      journal = {\araa},
         year = 2012,
        month = sep,
       volume = {50},
        pages = {531-608},
          doi = {10.1146/annurev-astro-081811-125610},
archivePrefix = {arXiv},
       eprint = {1204.3552},
 primaryClass = {astro-ph.GA},
       adsurl = {https://ui.adsabs.harvard.edu/abs/2012ARA&A..50..531K}
}

@ARTICLE{traina2024sfrd,
       author = {{Traina}, A. and {Gruppioni}, C. and {Delvecchio}, I. and {Calura}, F. and {Bisigello}, L. and {Feltre}, A. and {Magnelli}, B. and {Schinnerer}, E. and {Liu}, D. and {Adscheid}, S. and {Behiri}, M. and {Gentile}, F. and {Pozzi}, F. and {Talia}, M. and {Zamorani}, G. and {Algera}, H. and {Gillman}, S. and {Lambrides}, E. and {Symeonidis}, M.},
        title = "{A$^{3}$COSMOS: The infrared luminosity function and dust-obscured star formation rate density at 0.5 < z < 6}",
      journal = {\aap},
         year = 2024,
        month = jan,
       volume = {681},
          eid = {A118},
        pages = {A118},
          doi = {10.1051/0004-6361/202347048},
archivePrefix = {arXiv},
       eprint = {2309.15150},
 primaryClass = {astro-ph.GA},
       adsurl = {https://ui.adsabs.harvard.edu/abs/2024A&A...681A.118T}
}

@ARTICLE{schaye2015eagle,
       author = {{Schaye}, Joop and {Crain}, Robert A. and {Bower}, Richard G. and {Furlong}, Michelle and {Schaller}, Matthieu and {Theuns}, Tom and {Dalla Vecchia}, Claudio and {Frenk}, Carlos S. and {McCarthy}, I.~G. and {Helly}, John C. and {Jenkins}, Adrian and {Rosas-Guevara}, Y.~M. and {White}, Simon D.~M. and {Baes}, Maarten and {Booth}, C.~M. and {Camps}, Peter and {Navarro}, Julio F. and {Qu}, Yan and {Rahmati}, Alireza and {Sawala}, Till and {Thomas}, Peter A. and {Trayford}, James},
        title = "{The EAGLE project: simulating the evolution and assembly of galaxies and their environments}",
      journal = {\mnras},
         year = 2015,
        month = jan,
       volume = {446},
       number = {1},
        pages = {521-554},
          doi = {10.1093/mnras/stu2058},
archivePrefix = {arXiv},
       eprint = {1407.7040},
 primaryClass = {astro-ph.GA},
       adsurl = {https://ui.adsabs.harvard.edu/abs/2015MNRAS.446..521S}
}

@ARTICLE{adscheid2024a3cosmos,
       author = {{Adscheid}, Sylvia and {Magnelli}, Benjamin and {Liu}, Daizhong and {Bertoldi}, Frank and {Delvecchio}, Ivan and {Gruppioni}, Carlotta and {Schinnerer}, Eva and {Traina}, Alberto and {B{\'e}thermin}, Matthieu and {Gkogkou}, Athanasia},
        title = "{A$^{3}$COSMOS and A$^{3}$GOODSS: Continuum source catalogues and multi-band number counts}",
      journal = {\aap},
         year = 2024,
        month = may,
       volume = {685},
          eid = {A1},
        pages = {A1},
          doi = {10.1051/0004-6361/202348407},
archivePrefix = {arXiv},
       eprint = {2403.03125},
 primaryClass = {astro-ph.GA},
       adsurl = {https://ui.adsabs.harvard.edu/abs/2024A&A...685A...1A}
}

@ARTICLE{schmidt1959law,
       author = {{Schmidt}, Maarten},
        title = "{The Rate of Star Formation.}",
      journal = {\apj},
         year = 1959,
        month = mar,
       volume = {129},
        pages = {243},
          doi = {10.1086/146614},
       adsurl = {https://ui.adsabs.harvard.edu/abs/1959ApJ...129..243S}
}

@ARTICLE{pannella2015sfr,
       author = {{Pannella}, M. and {Elbaz}, D. and {Daddi}, E. and {Dickinson}, M. and {Hwang}, H.~S. and {Schreiber}, C. and {Strazzullo}, V. and {Aussel}, H. and {Bethermin}, M. and {Buat}, V. and {Charmandaris}, V. and {Cibinel}, A. and {Juneau}, S. and {Ivison}, R.~J. and {Le Borgne}, D. and {Le Floc'h}, E. and {Leiton}, R. and {Lin}, L. and {Magdis}, G. and {Morrison}, G.~E. and {Mullaney}, J. and {Onodera}, M. and {Renzini}, A. and {Salim}, S. and {Sargent}, M.~T. and {Scott}, D. and {Shu}, X. and {Wang}, T.},
        title = "{GOODS-Herschel: Star Formation, Dust Attenuation, and the FIR-radio Correlation on the Main Sequence of Star-forming Galaxies up to z ≃4}",
      journal = {\apj},
         year = 2015,
        month = jul,
       volume = {807},
       number = {2},
          eid = {141},
        pages = {141},
          doi = {10.1088/0004-637X/807/2/141},
archivePrefix = {arXiv},
       eprint = {1407.5072},
 primaryClass = {astro-ph.GA},
       adsurl = {https://ui.adsabs.harvard.edu/abs/2015ApJ...807..141P}
}

@ARTICLE{wuyts2011sfr,
       author = {{Wuyts}, Stijn and {F{\"o}rster Schreiber}, Natascha M. and {Lutz}, Dieter and {Nordon}, Raanan and {Berta}, Stefano and {Altieri}, Bruno and {Andreani}, Paola and {Aussel}, Herv{\'e} and {Bongiovanni}, Angel and {Cepa}, Jordi and {Cimatti}, Andrea and {Daddi}, Emanuele and {Elbaz}, David and {Genzel}, Reinhard and {Koekemoer}, Anton M. and {Magnelli}, Benjamin and {Maiolino}, Roberto and {McGrath}, Elizabeth J. and {P{\'e}rez Garc{\'\i}a}, Ana and {Poglitsch}, Albrecht and {Popesso}, Paola and {Pozzi}, Francesca and {Sanchez-Portal}, Miguel and {Sturm}, Eckhard and {Tacconi}, Linda and {Valtchanov}, Ivan},
        title = "{On Star Formation Rates and Star Formation Histories of Galaxies Out to z \raisebox{-0.5ex}\textasciitilde 3}",
      journal = {\apj},
         year = 2011,
        month = sep,
       volume = {738},
       number = {1},
          eid = {106},
        pages = {106},
          doi = {10.1088/0004-637X/738/1/106},
archivePrefix = {arXiv},
       eprint = {1106.5502},
 primaryClass = {astro-ph.CO},
       adsurl = {https://ui.adsabs.harvard.edu/abs/2011ApJ...738..106W}
}

@ARTICLE{fontanot2017highz,
       author = {{Fontanot}, Fabio and {Hirschmann}, Michaela and {De Lucia}, Gabriella},
        title = "{Strong Stellar-driven Outflows Shape the Evolution of Galaxies at Cosmic Dawn}",
      journal = {\apjl},
         year = 2017,
        month = jun,
       volume = {842},
       number = {2},
          eid = {L14},
        pages = {L14},
          doi = {10.3847/2041-8213/aa74bd},
archivePrefix = {arXiv},
       eprint = {1703.02983},
 primaryClass = {astro-ph.GA},
       adsurl = {https://ui.adsabs.harvard.edu/abs/2017ApJ...842L..14F}
}

@ARTICLE{springel2005DM,
       author = {{Springel}, Volker},
        title = "{The cosmological simulation code GADGET-2}",
      journal = {\mnras},
         year = 2005,
        month = dec,
       volume = {364},
       number = {4},
        pages = {1105-1134},
          doi = {10.1111/j.1365-2966.2005.09655.x},
archivePrefix = {arXiv},
       eprint = {astro-ph/0505010},
 primaryClass = {astro-ph},
       adsurl = {https://ui.adsabs.harvard.edu/abs/2005MNRAS.364.1105S}
}

@ARTICLE{crain2015eagle,
       author = {{Crain}, Robert A. and {Schaye}, Joop and {Bower}, Richard G. and {Furlong}, Michelle and {Schaller}, Matthieu and {Theuns}, Tom and {Dalla Vecchia}, Claudio and {Frenk}, Carlos S. and {McCarthy}, Ian G. and {Helly}, John C. and {Jenkins}, Adrian and {Rosas-Guevara}, Yetli M. and {White}, Simon D.~M. and {Trayford}, James W.},
        title = "{The EAGLE simulations of galaxy formation: calibration of subgrid physics and model variations}",
      journal = {\mnras},
         year = 2015,
        month = jun,
       volume = {450},
       number = {2},
        pages = {1937-1961},
          doi = {10.1093/mnras/stv725},
archivePrefix = {arXiv},
       eprint = {1501.01311},
 primaryClass = {astro-ph.GA},
       adsurl = {https://ui.adsabs.harvard.edu/abs/2015MNRAS.450.1937C}
}

@ARTICLE{lagos2018sam,
       author = {{Lagos}, Claudia del P. and {Tobar}, Rodrigo J. and {Robotham}, Aaron S.~G. and {Obreschkow}, Danail and {Mitchell}, Peter D. and {Power}, Chris and {Elahi}, Pascal J.},
        title = "{Shark: introducing an open source, free, and flexible semi-analytic model of galaxy formation}",
      journal = {\mnras},
         year = 2018,
        month = dec,
       volume = {481},
       number = {3},
        pages = {3573-3603},
          doi = {10.1093/mnras/sty2440},
archivePrefix = {arXiv},
       eprint = {1807.11180},
 primaryClass = {astro-ph.GA},
       adsurl = {https://ui.adsabs.harvard.edu/abs/2018MNRAS.481.3573L}
}

@ARTICLE{lacey2016sam,
       author = {{Lacey}, Cedric G. and {Baugh}, Carlton M. and {Frenk}, Carlos S. and {Benson}, Andrew J. and {Bower}, Richard G. and {Cole}, Shaun and {Gonzalez-Perez}, Violeta and {Helly}, John C. and {Lagos}, Claudia D.~P. and {Mitchell}, Peter D.},
        title = "{A unified multiwavelength model of galaxy formation}",
      journal = {\mnras},
         year = 2016,
        month = nov,
       volume = {462},
       number = {4},
        pages = {3854-3911},
          doi = {10.1093/mnras/stw1888},
archivePrefix = {arXiv},
       eprint = {1509.08473},
 primaryClass = {astro-ph.GA},
       adsurl = {https://ui.adsabs.harvard.edu/abs/2016MNRAS.462.3854L}
}

@ARTICLE{foremanmackey2013emcee,
       author = {{Foreman-Mackey}, Daniel and {Hogg}, David W. and {Lang}, Dustin and {Goodman}, Jonathan},
        title = "{emcee: The MCMC Hammer}",
      journal = {\pasp},
         year = 2013,
        month = mar,
       volume = {125},
       number = {925},
        pages = {306},
          doi = {10.1086/670067},
archivePrefix = {arXiv},
       eprint = {1202.3665},
 primaryClass = {astro-ph.IM},
       adsurl = {https://ui.adsabs.harvard.edu/abs/2013PASP..125..306F}
}

@ARTICLE{lawrence1986irlf,
       author = {{Lawrence}, A. and {Walker}, D. and {Rowan-Robinson}, M. and {Leech}, K.~J. and {Penston}, M.~V.},
        title = "{Studies of IRAS sources at high galactic latitudes - II. Results froma redshift survey at b>60 : distribution in depth, luminosity function, and physical nature of IRAS galaxies.}",
      journal = {\mnras},
         year = 1986,
        month = apr,
       volume = {219},
        pages = {687-701},
          doi = {10.1093/mnras/219.3.687},
       adsurl = {https://ui.adsabs.harvard.edu/abs/1986MNRAS.219..687L}
}

@ARTICLE{soifer1987irlf,
       author = {{Soifer}, B.~T. and {Sanders}, D.~B. and {Madore}, B.~F. and {Neugebauer}, G. and {Danielson}, G.~E. and {Elias}, J.~H. and {Lonsdale}, Carol J. and {Rice}, W.~L.},
        title = "{The IRAS Bright Galaxy Sample. II. The Sample and Luminosity Function}",
      journal = {\apj},
         year = 1987,
        month = sep,
       volume = {320},
        pages = {238},
          doi = {10.1086/165536},
       adsurl = {https://ui.adsabs.harvard.edu/abs/1987ApJ...320..238S}
}

@ARTICLE{sanders2003irlf,
       author = {{Sanders}, D.~B. and {Mazzarella}, J.~M. and {Kim}, D. -C. and {Surace}, J.~A. and {Soifer}, B.~T.},
        title = "{The IRAS Revised Bright Galaxy Sample}",
      journal = {\aj},
         year = 2003,
        month = oct,
       volume = {126},
       number = {4},
        pages = {1607-1664},
          doi = {10.1086/376841},
archivePrefix = {arXiv},
       eprint = {astro-ph/0306263},
 primaryClass = {astro-ph},
       adsurl = {https://ui.adsabs.harvard.edu/abs/2003AJ....126.1607S}
}

@ARTICLE{rush1993irlf,
       author = {{Rush}, Brian and {Malkan}, Matthew A. and {Spinoglio}, Luigi},
        title = "{The Extended 12 Micron Galaxy Sample}",
      journal = {\apjs},
         year = 1993,
        month = nov,
       volume = {89},
        pages = {1},
          doi = {10.1086/191837},
archivePrefix = {arXiv},
       eprint = {astro-ph/9306013},
 primaryClass = {astro-ph},
       adsurl = {https://ui.adsabs.harvard.edu/abs/1993ApJS...89....1R}
}

@ARTICLE{shupe1998irlf,
       author = {{Shupe}, David L. and {Fang}, Fan and {Hacking}, Perry B. and {Huchra}, John P.},
        title = "{The Local Luminosity Function at 25 Microns}",
      journal = {\apj},
         year = 1998,
        month = jul,
       volume = {501},
       number = {2},
        pages = {597-607},
          doi = {10.1086/305825},
archivePrefix = {arXiv},
       eprint = {astro-ph/9803149},
 primaryClass = {astro-ph},
       adsurl = {https://ui.adsabs.harvard.edu/abs/1998ApJ...501..597S}
}

@ARTICLE{wang2022molgas,
       author = {{Wang}, Tsan-Ming and {Magnelli}, Benjamin and {Schinnerer}, Eva and {Liu}, Daizhong and {Modak}, Ziad Aziz and {Jim{\'e}nez-Andrade}, Eric Faustino and {Karoumpis}, Christos and {Kokorev}, Vasily and {Bertoldi}, Frank},
        title = "{A$^{3}$COSMOS: A census on the molecular gas mass and extent of main-sequence galaxies across cosmic time}",
      journal = {\aap},
         year = 2022,
        month = apr,
       volume = {660},
          eid = {A142},
        pages = {A142},
          doi = {10.1051/0004-6361/202142299},
archivePrefix = {arXiv},
       eprint = {2201.12070},
 primaryClass = {astro-ph.GA},
       adsurl = {https://ui.adsabs.harvard.edu/abs/2022A&A...660A.142W}
}

@ARTICLE{katsianis2017z4,
       author = {{Katsianis}, A. and {Tescari}, E. and {Blanc}, G. and {Sargent}, M.},
        title = "{The evolution of the star formation rate function and cosmic star formation rate density of galaxies at z {\ensuremath{\sim}} 1-4}",
      journal = {\mnras},
         year = 2017,
        month = feb,
       volume = {464},
       number = {4},
        pages = {4977-4994},
          doi = {10.1093/mnras/stw2680},
archivePrefix = {arXiv},
       eprint = {1610.03441},
 primaryClass = {astro-ph.GA},
       adsurl = {https://ui.adsabs.harvard.edu/abs/2017MNRAS.464.4977K}
}

\begin{appendix}
\section{SFRF from SMF and MS}\label{app:cutouts}

\par In this Appendix, we describe the details of the sanity check performed using the SMF and the MS to obtain a SFRF. If we assume that galaxies follows a $M_{\star} \,-\, SFR$ correlation, it is possible to convert a measured SMF into a SFRF. For this test, we assumed the SMF by \citet{weaver2022cosmos2020}, derived from the COSMOS2020 sample, which roughly covers the redshift range between $z \sim 0.5$ and $z \sim 6$. To convert the stellar mass into SFR, we used two MSs, from \citet{speagle2014MS} and \citet{popesso2023MS}, which show a different behavior at large $M_{\star}$. In particular, \citet{speagle2014MS} find a linear trend for the MS, while the MS by \citet{popesso2023MS} exhibit the so-called bending in the high-mass end. However, this difference do not translate into a significant discrepancy in the final SFRF. In converting the stellar mass into $SFR$, we assumed that most of the galaxies follow the MS, with a certain dispersion, and a small percentage ($\sim 2\%$) is represented by starburts galaxies (defined here with a factor 5$\times$ higher than the MS in SFR). We find that, overall, the agreement with the observed data points and the SFRFs derived in this test is quite good, confirming the usability of our method.

\begin{figure*}[]
\centering
{\includegraphics[width=1.0\textwidth]{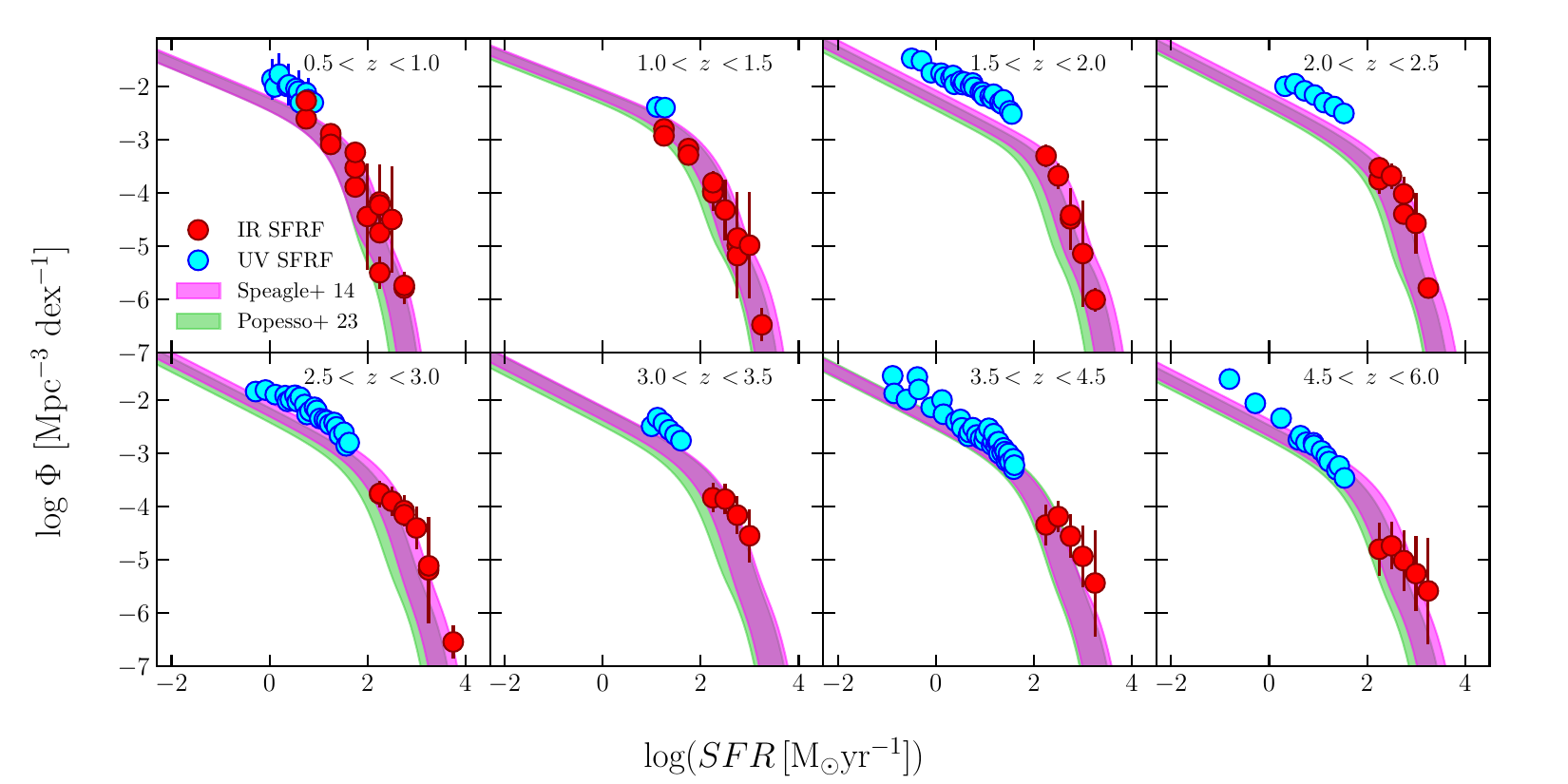}}
\caption{\small{IR (red circles) and UV (cyan circles) SFRFs (in complete SFR bins) compared to the SFRF obtained by converting the SMF using the MS. The magenta SFRF is obtained using the MS by \citet{speagle2014MS}, while the green area is from \citet{popesso2023MS}.}}
\label{fig:SMF_MS}
\end{figure*}

\section{Details on the models}\label{app:table_models}

We report in this Appendix (Tables \ref{tab:hydro} and \ref{tab:sam}) the main details, mostly on the dust and SF treatment, of the hydrodynamical simulations and SAMs used as a comparison in this work.

\begin{table*}[]
\renewcommand{\arraystretch}{1.5}
\caption{Key properties of the hydrodynamical simulations considered in this work, including star formation (SF) parameters.}
\begin{threeparttable}
\resizebox{\linewidth}{!}{%
\begin{tabular}{ccccccc}
\hline
             & \begin{tabular}[c]{@{}c@{}}Box-size\\ {[}Mpc/h{]}\end{tabular} 
             & \begin{tabular}[c]{@{}c@{}}Baryonic mass\\ resolution {[}M$_\odot${]}\end{tabular} 
             & \begin{tabular}[c]{@{}c@{}}Spatial\\ resolution {[}kpc{]}\end{tabular} 
             & Feedback 
             & Dust treatment
             & \begin{tabular}[c]{@{}c@{}}Star formation\\ ($\epsilon_{\rm SF}$, n$_{\rm H}$)\end{tabular} \\ \hline
EAGLE        & 100  & $\sim 1.8 \times 10^{6}$ & $\sim 0.7$ & SNe + AGN heating 
             & \begin{tabular}[c]{@{}c@{}}Not self-consistent; post-processed\\ dust models available\end{tabular}
             & \begin{tabular}[c]{@{}c@{}}$\epsilon_{\rm SF} \sim 0.02$, \\$n_{\rm H} = 0.1\, (Z/0.002)^{-0.64}\,$cm$^{-3}$\end{tabular} \\
IllustrisTNG & 110  & $\sim 1.4 \times 10^{6}$ & $\sim 1.0$ & \begin{tabular}[c]{@{}c@{}}SNe (winds) + \\ AGN (kinetic + thermal)\end{tabular} 
             & \begin{tabular}[c]{@{}c@{}}Not self-consistent; \\dust added in post-processing\end{tabular}
             & \begin{tabular}[c]{@{}c@{}}$\epsilon_{\rm SF} \sim 0.02$, \\$n_{\rm H} \sim 0.1$--$0.3\,$cm$^{-3}$ (metallicity-dependent)\end{tabular} \\
SIMBA        & 150  & $\sim 1.8 \times 10^{7}$ & $\sim 1.5$ & \begin{tabular}[c]{@{}c@{}}SNe + AGN\\ (radiative + jets)\end{tabular} 
             & \begin{tabular}[c]{@{}c@{}}Includes self-consistent dust\\ production and destruction\end{tabular}
             & \begin{tabular}[c]{@{}c@{}}$\epsilon_{\rm SF} \sim 0.02$, \\SF model based on molecular hydrogen fraction\end{tabular} \\ \hline
\end{tabular}}
\tablefoot{For each hydrodynamical simulation, the simulated volume, resolution, feedback implementation, dust treatment, and key star formation parameters are reported. Star formation efficiency ($\epsilon_{\rm SF}$) is per free-fall time and $n_{\rm H}$ is the hydrogen number density threshold for SF.}
\end{threeparttable}
\label{tab:hydro}
\end{table*}

\begin{table*}[]
\renewcommand{\arraystretch}{1.5}
\caption{Key properties of the semi-analytical models considered in this work, including star formation parameters.}
\begin{threeparttable}
\resizebox{\linewidth}{!}{%
\begin{tabular}{ccccc}
\hline
             & \begin{tabular}[c]{@{}c@{}}Halo catalog /\\ merger tree\end{tabular} 
             & Feedback 
             & Dust treatment
             & \begin{tabular}[c]{@{}c@{}}Star formation\\ ($\epsilon_{\rm SF}$, n$_{\rm H}$)\end{tabular} \\ \hline
GAEA         & \begin{tabular}[c]{@{}c@{}}Millennium \\ + rescaling\end{tabular} 
             & SNe + AGN outflows 
             & \begin{tabular}[c]{@{}c@{}}Includes self-consistent dust production\\ and chemical enrichment\end{tabular}
             & \begin{tabular}[c]{@{}c@{}}$\epsilon_{\rm SF} \sim 0.03$--$0.05$, Schmidt-Kennicutt type law,\\ $n_{\rm H} \sim 0.1$ cm$^{-3}$, $T \sim 10^4$ K\end{tabular} \\
L-GalP23     & Millennium 
             & \begin{tabular}[c]{@{}c@{}}SNe + AGN \\ (radio + quasar mode)\end{tabular} 
             & \begin{tabular}[c]{@{}c@{}}Dust treated via empirical/semi-analytic\\ prescriptions\end{tabular}
             & \begin{tabular}[c]{@{}c@{}}$\epsilon_{\rm SF} \sim 0.02$--$0.05$, Kennicutt-Schmidt law,\\ $n_{\rm H} \sim 0.1$ cm$^{-3}$, $T \sim 10^4$ K\end{tabular} \\ \hline
\end{tabular}}
\tablefoot{For each semi-analytical model, the halo catalog, feedback implementation, dust treatment, and key star formation parameters are reported. Star formation efficiency ($\epsilon_{\rm SF}$) is per free-fall time and $n_{\rm H}$ is the hydrogen number density threshold for SF.}
\end{threeparttable}
\label{tab:sam}
\end{table*}

\end{appendix}

\end{document}